\documentstyle[aps,12pt]{revtex}
\newcommand{\lettersize}{\baselineskip=0.5cm}
\newcommand{\dir}{FIGS1}
\input{epsf}
\newcommand{\fig}[3]
{
\begin{center}
     \noindent
     \unitlength=1mm
     \begin{picture}(#2,#3)
     \put(0,0){\leavevmode \epsfxsize=#2mm \epsffile{\dir/#1}}
     \end{picture}
   \noindent
\end{center}
}
\begin{document}
\setcounter{page}{1}

\baselineskip=1cm

\noindent
{\LARGE\bf
Short grafted chains: \\
Monte Carlo simulations of a model for monolayers of amphiphiles
}

\vspace{0.5cm}

\lettersize

\begin{center}
Christoph Stadler, Harald Lange, Friederike Schmid \\
{\em Institut f\"ur Physik, Universit\"at Mainz, D-55099 Mainz, Germany}
\end{center}


\begin{quote}
{\bf Abstract.}
We present Monte Carlo simulations of a coarse-grained model for 
Langmuir monolayers of amphiphile molecules on a polar substrate. The 
molecules are modelled as chains of Lennard-Jones beads, with one slightly
larger end bead confined in a planar surface. They are simulated in continuous 
space under conditions of constant pressure, using a simulation box of variable
size and shape. The model exhibits a disordered phase (corresponding to the 
liquid expanded phase), and various ordered phases (corresponding to the 
condensed phases) with different types of tilt. We calculate the phase 
diagrams and characterize the different phases and phase transitions. The 
effect of varying the chain stiffness is also discussed.
\end{quote}


\section{Introduction}
Monolayers of amphiphiles at surfaces (Langmuir monolayers) have attracted 
longstanding scientific interest for various
reasons\cite{mono-reviews,lb-books,kaganer-review}: 
Surface properties of materials can be modified and taylored by coating the 
surfaces with amphiphiles. Langmuir monolayers can be exploited to engineer 
thin film materials with well-defined  structures on a molecular level. 
On the other hand, lipid monolayers on water are experimentally fairly 
accessible model systems for biological membranes. Last not least, Langmuir 
monolayers are experimental realizations of two-dimensional systems, which 
allow to study ordering phenomena in low dimensions.

Experimentally, Langmuir monolayers have been investigated for a long time
by measurements of pressure-area isotherms\cite{mono-reviews}. More recently, 
a number of powerful microscopy techniques have been developed, such as
fluorescence microscopy and Brewster angle microscopy,
which have provided insight into the mesoscopic structures in monolayers.
The emerging pictures for monolayers on water is qualitatively similar
for phospholipids, long chain alcohols and esters: At low surface coverage,
the molecules hardly interact with each other and build the two-dimensional 
equivalent of a ``gas''. Upon compression, a first order transition to a 
fluid-like ``liquid expanded'' (LE) phase is encountered, followed at
even higher surface coverage by a second discontinuous transition into a 
``liquid condensed''(LC) area. The transition from liquid expanded to liquid 
condensed has an important equivalent in bilayers, the ``main transition'', 
which may be biologically relevant, since it takes place at temperatures
close to the body temperature for some of the common phospholipids. The 
condensed region contains a variety of different phases, characterized by 
different types of ordering, {\em i.e.}, collective tilt order of the
hydrocarbon chains, orientational order of the backbones of the chains, and
crystalline positional order. A generic phase diagram for 
fatty acid monolayers is shown in Fig. \ref{fig1}\cite{kaganer-review}. 
The lowest density phases which coexist with the LE phase are typically 
hexatic rotator phases, {\em i.e.}, the backbones rotate freely around, and 
positional correlations decay exponentially, but the directions of
nearest neighbors are nevertheless well-defined.

Theoretical treatments of Langmuir monolayers have followed three different
lines. On the one hand, phenomenological descriptions of the 
different condensed phases in terms of Landau expansions in the 
characteristic order parameters\cite{selinger,kaganer} have offered
valuable insight into the nature and the interrelations of 
different phase transitions on a very general level. On the
other hand, Molecular dynamics simulations of atomically realistic
models have complemented experiments and provided structural
information on quantities, which are hard to access 
experimentally\cite{kox,klein,alper,rice,karaborni}. 
These two approaches are in a sense antipodal: 
Whereas phenomenological treatments focus on universal properties and
make little or no contact to the microscopic structure of the systems,  
atomically realistic models seek to imitate nature as faithfully
as possible, and to reach quantitative agreement. Hence they account for
many more details than are actually needed to produce a certain phase
behavior, rely heavily on the availability of good force fields,
and their study is computationally costly. 

As a third line of approach, idealized microscopic models are constructed
which incorporate only a few properties of a material, believed to be 
essential for a given behavior. Thus they bridge between phenomenological and 
realistic models, and relate microscopic and macroscopic quantities
in a qualitative and semi-quantitative way.

The question, which features of amphiphiles are essential in Langmuir
monolayers, can of course not be answered universally. It depends on
the region in phase space one wishes to study.  Attractive interactions 
between the amphiphiles are important for most phase transitions. 
As long as one studies condensed phases, it is often sufficient to model the 
amphiphiles as anisotropic stiff objects. Grafted rigid rods exhibit tilt 
transitions\cite{binder-tilt,tilt,opps}, molecules with non-circular 
cross-sections show rotator transitions\cite{swanson}. For the transition 
between the liquid condensed and the liquid expanded phase, however, the 
conformational degrees of freedom of the chains play a crucial 
role\cite{rice,barton,me1}. 
They have been incorporated in a heuristic way as ``internal degeneracies'' 
in Ising-type two-dimensional lattice models for monolayers and bilayers, 
{\em e.g.}, in the Pink model\cite{pink,bernd-review}.
The interdependence of chain conformations and effective chain interactions 
has to be put in by hand in this approach, and a large number of input 
parameters is required. Models which aim to study more directly the interplay of
chain conformations and phase behavior have to retain the chain character
of the amphiphiles explicitly.

A suitable idealized model for Langmuir monolayers thus represents 
the amphiphiles by flexible chains of mutually attracting monomers, which are 
grafted to a surface at one end (``head''). Such models have been formulated 
on the lattice\cite{szleifer,rice-lat,kolinski,stettin,haas1} and in 
continuous space\cite{haas2,haas3,christoph1,harald-dipl,christoph-diss}. 

Lattice models can be simulated more efficiently than off-lattice models, 
yet they can produce rather awkward lattice effects especially when 
orientational order (tilt order) comes into play\cite{haas1}.
An off-lattice bead-spring model of Lennard-Jones beads has been studied by
Haas {\em et al}\cite{haas2,haas3} and by us\cite{christoph1} under constant 
volume and constant pressure conditions. It was found to 
display a tilted and an untilted phase, in which the chains are basically
arranged on a (possibly distorted) hexagonal lattice, and a ``fluidized'' 
phase which is reminiscent of the liquid expanded phase. Hence it seems a 
promising candidate for a minimal model, which contains only the basic 
elements responsible for the main transition in Langmuir monolayers. 
Nevertheless, no systematic study of the phase behavior has been presented 
so far. 

This is the objective of the present paper. We have performed 
Monte Carlo simulations of a bead-spring model very similar to the 
one used by Haas {\em et al}. The models only differ in the treatment
of the heads: Whereas the head beads in Haas {\em et al}'s version are 
identical with the chain beads, our heads are slightly larger.
We chose this variant in order to ensure that the dominant reason for chain 
tilting in our model is similar to the most common one in nature: Tilt is
induced by the mismatch of head and tail size.
In the model of Haas {\em et al}, the chains tilt, because they 
can then ``hook'' into each other and thus pack more efficiently. The details
of the tilt order (tilt angle, tilt direction etc.) result from a complicated 
interplay between monomer packing and chain stretching\cite{haas3}, which is 
highly model dependent and has probably little to do with the factors which
influence the tilt in real monolayers.
On simple geometrical grounds, two of us have argued earlier that the
direction of tilt depends on the size of the head groups\cite{me2}.
There is also experimental evidence for such a connection\cite{teer}.
With our choice of the head size, we ensure that the model exhibits 
two different tilted phases at zero temperature, a low-pressure one
with tilt towards nearest neighbors, and a higher-pressure one with
tilt towards next nearest neighbors.

Our paper is organized as follows: In the next section, we specify the
model and comment on some aspects of the simulation techniques and the
data analysis. The results are presented in section 3: We characterize the 
phases and phase transitions, show the phase diagrams, and 
discuss the effect of the chain stiffness. We summarize and conclude in
section 4.

\section{Model and technical details}

Following Haas {\em et al}\cite{haas2,haas3}, we model the amphiphiles as 
chains of beads, which are connected by springs of length $d$ subject to the 
spring potential
\begin{equation}
\label{fene}
V_{S}(d) = \left\{ \begin{array}{l c r}
- \frac{k_{S}}{2} d_{S}^2 \ln\Big( 1 - (d-d_0)^2/d_{S}{}^2 \Big)
& \mbox{for} & |d-d_0|<d_{S} \\
\infty & \mbox{for} & |d-d_0| > d_{S}
\end{array} \right. 
\end{equation}
This so-called ``Finite extension nonlinear elastic'' potential (FENE)
is basically harmonic at $d \approx d_0$ and has a logarithmic cutoff
at $d=d_0 \pm d_{S}$. Furthermore, we impose a stiffness potential
\begin{equation}
\label{ba}
V_{A} = k_{A} \cdot (1- \cos \theta)
\end{equation}
on the angle $\theta$ between subsequent springs. The stiffness potential
favors angles $\theta=0$, {\em i.e.}, straight chains. Beads are not 
allowed to enter the half space $z<0$; moreover, one end bead of each
chain (the ``head'') is confined to remain within the plane $z=0$.
Thus we assume a very strong binding force between the hydrophilic
head group and the water surface, and the latter is approximated by
a perfectly sharp and flat interface.
Tail beads interact {\em via} a truncated Lennard-Jones potential
\begin{equation}
\label{vlj}
V_{LJ}(r) = \left\{ \begin{array}{lcr}
\epsilon \cdot 
\Big( \big(\sigma/r\big)^{12} - 2 \big(\sigma/r\big)^6 + v_c \Big) &
\mbox{for} & r \le 2 \sigma \\
0 & \mbox{for} & r > 2 \sigma
\end{array} \right. ,
\end{equation}
where $v_c = 127/4096 \approx 0.031$ is chosen such that $V_{LJ}(r)$
is continuous at $r= 2\sigma$. The interactions between head beads
are purely repulsive,
\begin{equation}
\label{vh}
V_{H}(r) = \left\{ \begin{array}{lcr}
\epsilon_H \cdot 
\Big( \big(\sigma_H/r\big)^{12} - 2 \big(\sigma_H/r\big)^6 + 1 \Big) &
\mbox{for} & r \le  \sigma_H \\
0 & \mbox{for} & r > \sigma_H
\end{array} \right. .
\end{equation}
The attractive part here has been cut off for reasons of computational 
efficiency. Note that the head size $\sigma_H$ differs
from the tail bead size $\sigma$. Head and tail beads interact with
a repulsive potential of the form (\ref{vh}), in which $\sigma_H$ is replaced
by $(\sigma_H + \sigma)/2$.

The parameters $\epsilon$ and $\sigma$ define the units of energy and length.
To complete the definition of the model, we have to specify the remaining 
parameters $d_0$, $d_S$, $k_S$, $k_A$, $\epsilon_H$, and $\sigma_H$:
Our choice was motivated by the idea that one bead should represent
roughly two CH${}_2$ groups in an actual alkane chain. Comparing a
straight model chain with an ideal all-trans state hydrocarbon chain, with
realistic potential parameters of united-atom potentials taken from
the literature ({\em e.g.}, from Ref. \cite{rigby}), one finds that the bond 
length $d_0$ should be approximately 0.7 times the chain diameter, 
$d_0=0.7 \sigma$. The identification also allows for a rough estimate 
of the absolute values of $\sigma$ and $\epsilon$: $\sigma \approx 3.8 \AA$ and
$\epsilon \approx 240 k_B K$, where $k_B$ is the Boltzmann constant.
These values should of course not be taken too literally, since the
model is much too simple to allow for quantitative comparisons
with experimental systems.

The spring constant $k_S$ was chosen very strong, $k_S = 100 \epsilon$,
such that the lengths of the springs are approximately constant
at all temperatures of interest. The value of the cutoff $d_S$ then has little
influence on the properties of the system; we use $d_S = 0.2 \sigma$. 
The stiffness constant $k_A$ can be estimated by adjusting the average
$\langle \cos \theta \rangle$ of a single free chain in our model at
a given temperature to the corresponding value in a single free alkane
chain. Such an estimate would yield $k_A \approx 5 \epsilon$ at room
temperature. Haas {\em et al}\cite{haas2,haas3} have used $k_A=10 \epsilon$.
Here, we have mostly used the same value ($k_A = 10 \epsilon$) in order
to be consistent with their work. For the reasons mentioned in the 
introduction, the size of the head beads was taken to be 
$\sigma_H = 1.1 \sigma$. The influence of the head size on the phase behavior 
shall be discussed in detail elsewhere\cite{christoph-diss,christoph2}. 
The prefactor $\epsilon_H$ was chosen $\epsilon_H=\epsilon$.

The simulations were performed at constant spreading pressure in a 
simulation box of variable size and shape. More specifically, we study
$n$ chains of length $N$ on a parallelogram with side length $L_x$ and $L_y$ 
and angle $\alpha$. Periodic boundary conditions were applied in these
two directions, and free boundary conditions in the third. Our Monte
Carlo moves include:
\begin{itemize}
\item Attempts to displace single beads
\item Attempts to vary $L_x$, $L_y$ or $\alpha$, {\em i.e.}, to rescale
  all coordinates such the the configuration is stretched or squeezed
  in one direction, or sheared (``volume moves'')
\end{itemize}
The trial moves are accepted or rejected according to a standard
Metropolis prescription with the effective Hamiltonian\cite{allen-tildesley}
\begin{equation}
\label{heff1}
H = E + \Pi A - n N T \ln(A),
\end{equation}
where $E$ is the internal energy, $\Pi$ the applied spreading pressure, and
$A = L_x L_y \sin \alpha$ the area of the simulation box. We have also
implemented collective moves, in which chains were displaced as a whole,
and volume moves, in which only the coordinates of the head beads are rescaled,
but inner molecular distances and angles are kept constant.
The Hamiltonian (\ref{heff1}) then has to be replaced by
\begin{equation}
\label{heff2}
H = E + \Pi A - n T \ln(A).
\end{equation}
Unfortunately, these collective moves did not reduce the time needed to 
generate uncorrelated configurations significantly.
Similarly, we have implemented continuous configurational biased Monte
Carlo moves\cite{cb}, but found that they brought no improvement in our
particular system. 

In order to check that no internal stress is present in our simulations, we 
have determined the internal pressure tensor
\begin{equation}
{\Pi}^{int}_{\alpha \beta} = \frac{1}{A} 
\langle \sum_{i=1}^{n N} r_{i \alpha} {F}_{i \beta} \rangle
 + \frac{N k_B T}{A} \delta_{\alpha \beta},
\end{equation}
where the sum $i$ runs over all monomers, $\alpha, \beta$ over the $x$ and
$y$ coordinate, $\vec{F}_i$ denotes the force acting on monomer $i$, and 
$\delta_{\alpha \beta}$ is the unit matrix. According to the virial
theorem, $\Pi^{int}_{\alpha \beta}$ should be diagonal and identical 
to $\Pi \delta_{\alpha \beta}$ at mechanical equilibrium. This was the case
in our simulations, if we used a simulation box of variable shape.
In simulation runs with a rectangular box, we sometimes obtained nonzero 
off-diagonal elements $\Pi_{x y}$ in the tilted phases.

We will present results for $n=144$ chains of length $N=7$. The average
decorrelation time lies between 200 and 1000 Monte Carlo steps (MCS), where
one MCS consists of $Nn=1008$ attempts of monomer moves,
and one attempt to rescale $L_x$, $L_y$, and $\alpha$. In general, the
systems were equilibrated during 70.000 MCS, and data were then collected
from every 500st configuration over a period of at least 200.000 MCS.

The simulations were supplemented by a low temperature analysis. 
The zero temperature ground state was determined by minimization of the
enthalpy (\ref{heff1}). A harmonic expansion was then performed in
order to determine the free energy $G$ at some given low (nonzero) temperature 
$T_0$. Given this reference value, one can calculate the free energy at other 
temperatures and pressures from simulations by means of a thermodynamic 
integration 
\begin{equation}
G(\Pi,T) = G(\Pi_0,T_0) + k_B T \int_{\Gamma} \Big\{
d \Pi' \frac{A}{k_B T'} - d T' \frac{H}{k_B T'^2} \Big\},
\end{equation}
as long as the path $\Gamma$ from $(\Pi_0,T_0)$ to $(\Pi,T)$ does not cross a 
first order phase transition. By comparing the free energies of different
states, we have localized the transition points between phases at low 
temperatures where hysteresis effects were strong.

\section{Results}

Figure \ref{fig2} shows temperature-area isobars for a selection of low 
pressures (a) and high pressures (b). One clearly observes a jump in the area 
per molecule, which moves to higher temperatures as the pressure increases. 
At high pressures, one discernes in addition a kink at low temperatures, 
indicating the presence of a second phase transition. 

The phases can be characterized by the typical features of the pair correlation 
functions (Figs. \ref{fig3} and \ref{fig4}) and structure functions 
(Figs. \ref{fig5} and \ref{fig6}). For example, the two dimensional correlation 
functions in the intermediate temperature state at high pressures are precisely 
those of a hexagonal lattice. Fig. \ref{fig3} shows pair correlation functions 
for the head groups, the projection of center of gravity of the chains onto 
the $xy$ plane, and the points where the chains pass through the plane at 
$z=2 \sigma$ above the surface at pressure $\Pi=100 \epsilon/\sigma^2$ and 
temperature $T=1 \epsilon/k_B$, which is slightly above the first phase 
transition. The three curves do not differ from each other qualitatively,
and the position and relative heights of the peaks are consistent with
those of a hexagonal structure. At temperatures below the first phase
transition or at lower pressures, each of the peaks splits up in two. 
This indicates that the hexagonal lattice is distorted in one of the high 
symmetry directions, either the nearest neighbor or the next nearest neighbor 
direction (for intermediate directions the peaks would split up in three). 
An example is shown in Fig. \ref{fig4} (see the curves for the lowest 
temperature $T=0.1 \epsilon/k_B$). From the large height difference of the twin 
peaks, one can infer that the lattice is stretched in the direction of nearest 
neighbors in this specific case. The direct inspection of configuration 
snapshots reveals, not surprisingly, that the lattice distortion goes
along with a collective tilt of the chains in the direction of the distortion.
At low pressures ($\Pi \stackrel{<}{\sim} 10 \epsilon/\sigma^2$), 
the hexagonal lattice in the tilted phases is stretched by roughly 10~\%.

With increasing temperature, the structure of the correlation functions is 
gradually lost. Slightly below the phase transition, the correlation functions 
of the head lattice are fluid-like, with peaks of monotonically decreasing 
height for the first, second and third coordination shell. They do not change 
qualitatively as the phase transition is crossed (Fig. \ref{fig4} (a)). 
In contrast, the correlation function for the projections of the center of 
gravity still shows some solid-like structure right below the phase transition,
and loses almost every structure right above the phase transition 
(Fig. \ref{fig4} (b)). In the high temperature state, the head positions are 
much more correlated than the chain positions. We conclude that the phase 
transition associated to the area jump is a melting transition, and that
it is driven by the chains. The chains maintain the order below the transition, 
and promote the disorder above the transition. This is consistent with results 
from molecular dynamics simulation by Karaborni and Toxvaerd\cite{karaborni}
of a realistic model.

The structure function is defined by
\begin{equation}
S(\vec{q}) = \frac{1}{n N} 
\bigg| \sum_{j=1}^{n N} \exp(i \vec{q} \vec{r}_j )\bigg|^2,
\end{equation}
where the sum runs over all monomers in the system. Note that in a
finite simulation box with periodic boundary conditions, $S(\vec{q})$ for
a specific configuration is only defined for vectors $\vec{q}$ whose projection 
on the $xy$ plane are sums of integer multiples of the basis vectors
\begin{displaymath}
\vec{b}_x = \frac{2 \pi}{L_x} {1 \choose -1/\tan(\alpha)} 
\qquad \mbox{and} \qquad
\vec{b}_y = \frac{2 \pi}{L_y} {0 \choose 1/\sin(\alpha)}.
\end{displaymath}
However, the dimensions of the box fluctuate in our simulations, hence the
basis vectors fluctuate as well. In order to overcome this problem, we
have laid a fine-meshed grid on the $xy$ plane and summed up all the 
contributions to $S(\vec{q})$ within a mesh. Fig. \ref{fig5} (a) and (b)
shows the resulting structure factors in the plane of $q_z=0$ for an
disordered state (a) and an untilted ordered state (b). The structure
factor of the disordered state is isotropic and shows the usual features
of a fluid structure factor. In the untilted ordered state, one finds the
Bragg rods of the hexagonal lattice. They are sharply peaked in the $xy$ 
plane, but have a considerable width in the $z$ direction, thus the term 
``rods''. In the tilted ordered state, the plane of maxima tilts such that
it stays perpendicular to the long axis of the chains\cite{kaganer-review}. 
Thus the peaks belonging to $\vec{q}$ vectors which are not perpendicular to 
the tilt direction move out of the $q_z=0$ plane. This is illustrated in 
Fig \ref{fig6} for a state with tilt towards next nearest neighbors. 
The internal structure of the rods in the $z$-direction reflects the 
structure of the monolayer. For example, the width of the rods is
inversely proportional to the width of the layer, and every rod is surrounded
by a multitude of weak ``satellite maxima'' which are caused by the sharp 
steps in the density profile at $z=0$ and at the outer surface.
After six low satellite maxima, another strong peak is
found, reaching a height comparable to that of the main peak. These peaks 
reflect the ``periodic'' arrangement of monomers {\em within} a chain. 
They are found at distances of approximately 
$\Delta q_z \approx 2 \pi/ d_0 \cos \theta$ and integer multiples from the
main peak, where $d_0$ is the favored distance between monomers 
(see eqn. (\ref{fene})). Their appearance is a very specific property 
of our simulation model, and not interesting from a general point of view.
Hence they shall not be studied any further.

In order to quantify our findings, we have analyzed a number of suitable order 
parameters. For example, we determine the hexagonal order parameter of 
two dimensional melting
\begin{equation}
\label{psi}
\Psi_6 = \bigg\langle \bigg| \frac{1}{6n}
\sum_{j=1}^n \sum_{k=1}^6 \exp(i 6 \phi_{jk}) \bigg|^2 \Big\rangle.
\end{equation}
Here the first sum $j$ runs over all heads of the systems,
the second $k$ over the six nearest neighbors of $j$, and $\phi_{jk}$ is the 
angle between the vector connecting the two heads and an arbitrary 
reference axis. The quantity $\Psi_6$ thus measures the orientational
long range order of nearest neighbor directions. It is nonzero in the
hexagonal (quasi)crystalline phase and in the hexatic phase.
As an order parameter which describes the collective tilt of molecules,
we have computed
\begin{equation}
R_{xy} = \sqrt{\langle [x]^2 + [y]^2 \rangle },
\end{equation}
which corresponds to the length of the average projection of the
head-to-end vector of the chains on the $xy$ plane. Here $[x]$ and $[y]$ 
denote the $x$ and $y$ component of the head-to-end vector, averaged over
the chains of a configuration, and $\langle \cdot \rangle$ denotes
the thermal average over all configurations. The quantitity $R_{xy}$ 
is nonzero in phases which break the azimuthal symmetry, {\em i.e.}.
phases with collective tilt, and zero otherwise. Note that the average
tilt angle $\theta$ between the head-to-end vector of the chains
and the surface normal is always nonzero.

The quantities $\Psi_6$ and $R_{xy}$ are shown as a function of temperature for 
various pressures in Figs. \ref{fig7} and \ref{fig8}. The area jump in the 
isobars goes along with a drop to almost zero of the melting order parameter 
$\Psi_6$. This substantiates our earlier speculation that the
transition corresponds to a melting transition. Furthermore, we infer from the 
decrease of $R_{xy}$ with the temperature, that there is also a tilting 
transition from a collectively tilted phase at small temperatures, to an 
untilted phase at high temperature. The melting transition and the tilting 
transition occur simultaneously at low pressures, and decouple from each other 
at high pressures. The tilting transition then precedes the melting transition 
and seems to be continuous. 

At our small system size, it is not possible to decide whether the ordered
phase is crystalline or hexatic. Moreover, we are not able to establish
unambiguously the order of the melting transition. These are two closely 
related issues of high interest. Even in much simpler two dimensional 
systems (hard disks, Lennard-Jones disks), the question whether they
melt discontinuously in one stage, or continuously {\em via} a hexatic 
phase\cite{halperin-nelson} in two stages, is still a matter of debate.
The transition from a hexatic to a fluid phase is usually believed to be 
continuous. In the case of amphiphile monolayers, however, we have argued 
that it can be driven first order, as an effect of the interplay between 
chain entropy and chain packing\cite{me1}.
We have already noted that the melting transition in our system is mainly 
driven by the chains, which enhances the likelihood of such a scenario.
The transition may also be discontinuous at low and intermediate pressures, 
and continuous at high pressures. The pronounced jumps observed in our 
simulations seem to indicate a line of discontinuous transitions; 
on the other hand, we have not encountered significant hysteresis effects 
except at very low pressure, $\Pi=1$. Simulations of much larger systems and 
a thorough finite-size analysis would presumably be necessary to distinguish 
between first order and continuous transitions.

It is instructive to also consider the distribution of tilt angles $\theta$. 
Let us first look at the average $\langle \theta \rangle$ (Fig. \ref{fig9}). 
In the low pressure regime, where the melting and the tilting phase transition 
coincide, it drops down at the transition and then rises slowly with 
temperature. At higher pressures, where the two transitions decouple, it first 
decreases with temperature until the tilting transition is passed, then stays 
low in the temperature region of the untilted ordered phase, but jumps to a 
higher value at the melting transition. The jump is related to the jump in the 
area per molecule at that transition: the molecules have more space to lie down.
The average tilt angle is coupled to the molecular area $A/n$ by the
requirement that the bead density in the monolayer should not vary much, i.e., 
the total volume occupied by the monolayer is close to constant. In the 
condensed region, where the chains are mostly straight and aligned, this 
implies that the quantity $A \cos (\theta)/n$ is approximately constant and 
equal to $a_c$, the area per molecule in the untilted high pressure phase. 
Such a dependence has indeed been reported experimentally\cite{shih}. 
Similarly, we find here that the product of $A$ and 
$\cos(\langle \theta \rangle)$ 
depends much less on the temperature and pressure than the area per 
molecule $A/n$ itself (Figure \ref{fig10}). In particular, its value right
below the melting transition is found to be $a_c \sim 0.985 \sigma^2$ at all
pressures except for the very highest, $\Pi = 100 \epsilon/\sigma^2$, 
regardless of whether the condensed phase is tilted or untilted.
Hence the {\em volume} density in the monolayer seems to trigger the
melting transition rather than the area density -- which corroborates our
earlier assertion that the melting transition is driven by the chains.

Figure \ref{fig11} shows the histogram of the tilt angle 
$P(\theta)/\sin\theta$ at pressure $\Pi=50$ for different temperatures. Below 
the tilting transition, $P(\theta)/ \sin \theta$ has a clear maximum.
As the temperature is increased, the maximum moves down towards lower values
$\theta$. At the tilting transition, it merges into $\theta = 0$. From 
there on, it becomes broader, which explains the increase of 
$\langle \theta \rangle$ at higher temperatures.

Finally, we turn to the discussion of the direction of the tilt. It can be 
determined from a histogram of the angle between the momentary tilt direction 
and the bonds connecting nearest neighbors. If the tilt angle is well-defined, 
this histogram should have six peaks, and their positions 
indicate the direction of tilt. At low temperatures $T \stackrel{<}{\sim} 0.1$, 
we find two phases with well-defined tilt directions towards nearest neighbors
and next nearest neighbors. The transition between them is strongly first
order, and the thermodynamic integration methods described in the previous
section had to be used to locate the transition points. At higher
temperatures, the transition washes out, and in some regions of phase space
it is hard to determine whether the tilt direction is at all locked to the
underlying hexagonal head lattice. In order to quantify the ``locking'', we
define an order parameter $\Phi_6$, which is very similar to the hexagonal
order parameter $\Psi_6$ (eqn. (\ref{psi})).
\begin{equation}
\label{phi}
\Phi_6 = \bigg| \bigg\langle \frac{1}{6n}
\sum_{j=1}^n \sum_{k=1}^6 \exp(i 6 \phi'_{jk}) \bigg\rangle \bigg|^2.
\end{equation}
The notation corresponds to that in eqn. (\ref{psi}), except that $\phi_{jk}'$ 
is now the angle to the average tilt direction in the current configuration
rather than simply that to an arbitrary reference axis. The crucial difference 
to the definition of $\psi_6$ lies in the detail that the sequence of 
$\langle . \rangle$ and $| . |$ has been interchanged. The parameter
$\Phi_6$ is nonzero if the tilt direction is locked to the nearest neighbor,
next nearest neighbor, or to an intermediate direction. However, it would still
be zero in a special case of locked state, where the tilt jumps between nearest 
and next nearest neighbors. In order to distinguish such a state from one
where the tilt direction is really oblivious to the hexagonal lattice, we have 
also evaluated the related parameter $\Phi_{12}$
\begin{equation}
\Phi_{12} = \bigg| \bigg\langle \frac{1}{6n}
\sum_{j=1}^n \sum_{k=1}^6 \exp(i 12 \phi'_{jk}) \bigg\rangle \bigg|^2.
\end{equation}
The parameter $\Phi_6$ and $\Phi_{12}$ are shown in Fig. \ref{fig12} for fixed 
temperature $T=0.5 \epsilon/k_B$ as a function of pressure. At this temperature,
the monolayer is tilted at all pressures shown. Fig. \ref{fig7} demonstrates 
that the tilt direction is locked to the hexagonal lattice at low pressures, but 
apparently unlocks at $\Pi = 40 \epsilon/\sigma^2$. That unlocked phases should 
exist in tilted {\em hexatic} liquid crystal films has been claimed by Selinger 
and Nelson\cite{selinger2}. In crystalline phases, they are supposedly 
suppressed by the elastic interactions. Since our systems are too small to 
allow for a distinction between hexatic and crystalline order, they are 
obviously also too small to allow to decide whether the unlocked state is real 
or a finite-size artefact.

In order to study the role of the chain flexibility, we have also
performed a few shorter simulation runs (35.000 MCS) of systems with stiffer
chains\cite{harald-dipl}. To this end, the stiffness constant $k_A$
(cf. eqn. (\ref{ba})) was increased by a factor of ten,
$k_A=100 \epsilon$. The area per molecule $A/n$, the melting order parameter 
$\Psi_6$ and the order parameter of collective tilt $R_{xy}$ for these systems 
are shown as a function of temperature for three different pressures
$\Pi = 10, 30$ and $40 \epsilon/\sigma^2$ in Fig. \ref{fig13}. 
Up to the highest pressure $\Pi = 40 \epsilon/\sigma^2$, the melting
transition and the tilting transition are coupled. Moreover, the melting
transition is shifted to much higher temperatures. This demonstrates once
more that the melting transition in the system is basically driven
by the chains.

Our results for flexible chains are summarized in the phase diagrams 
Fig. \ref{fig14} and Fig. \ref{fig15}. 
We find at least four phases: the disordered fluid,
an untilted ordered phase, two tilted ordered phases with tilt towards
nearest neighbors and next nearest neighbors, and possibly an unlocked
tilted phase. The areas per molecule of the two locked tilted phases
are almost equal at the transition, even at low temperatures where the 
latter is strongly first order. At higher temperatures, the transition 
is so washed out that it cannot be located any more. The transition between
the tilted and the untilted ordered phase seems continuous.  Between
the tilted ordered phase and the disordered phase, it is presumably
first order. The order of the transition between the untilted ordered
phase and the disordered phase could not be determined, as discussed
above.  It should be stressed that none of our assertions on the order
of the transitions has been corroborated by a finite size analysis,
hence they should be regarded with caution.

At surface areas per molecules smaller than $A \approx 0.8 \sigma^2$,
{\em i.e.}, at high pressures and low temperatures, the chains are
squeezed together so closely that they form ``rippled'' structures
where the beads of chains in neighbor rows are displaced with respect
to each other in the $z$ direction. This effect is clearly an artefact
of our model and has not been investigated in detail, nor included in
the phase diagram Fig. \ref{fig13}. In the limit of vanishing pressure,
on the other hand, the system has to assume a gas phase at all
temperatures for entropic reasons. The transition between the gas phase
and the condensed phase is subject to strong hysteresis effects at low
temperatures. Nevertheless, we have been able to determine the area per
molecule of the coexisting condensed state without too much
computational effort on the basis of the following consideration: An
upper limit is given by the area per molecule of the metastable
condensed state at zero pressure, which does not decay within the
simulation time at temperatures below $T = 1.35 \epsilon/k_B$. A lower
limit is provided by the area per molecule at the smallest pressure for
which the transition temperature from the ordered to the disordered
state has been determined, in our case $\Pi = 1 \epsilon/\sigma^2$.
Since the areas per molecule do not depend strongly on the pressure in
the condensed state, the coexistence line can thus be located fairly
accurately (see Fig. \ref{fig14}).

Within the region of the disordered fluid, we have not found evidence
for an additional liquid/gas transition. Such a transition would be
expected at areas per molecule much larger than $\sim 3 \sigma^2$
(where the critical point is found in two dimensional Lennard-Jones
fluids \cite{nigel}), and correspondingly low surface pressures. We have
spent some time searching for it, varying the temperature at very low
pressure $\Pi=0.05 \epsilon/\sigma^2$, and driving the pressure to zero
at the temperature $T=1.45 \epsilon/k_B$\cite{footnote}. In a region
around ($\Pi=0.05 \epsilon/\sigma^2, T \approx 1.7 \epsilon/k_B$) or
($\Pi \approx 0.04 \epsilon/\sigma^2, T = 1.45 \epsilon/k_B$), the area per
molecule varied rapidly, and strong density fluctuations were encountered.
This suggests that liquid-gas critical point may be nearby. However, we have 
not been able to locate it so far. It may be hidden in the coexistence
region.

\section{Conclusions}

To summarize, we have studied in detail the phase behavior of a model
of grafted Lennard-Jones chains, which is of interest as a ``minimal''
model for amphiphile monolayers. The model was found to show an impressive 
variety of phases, and its analysis gives useful insight into the mechanisms 
which drive some of the phase transitions in amphiphilic layers. In
particular, it exhibits a disordered phase, an untilted ordered phase, and a 
number of tilted ordered phases, which are also found experimentally in 
Langmuir monolayers. The sequence of tilting transitions with increasing 
pressure (tilt towards nearest neighbors, tilt towards next nearest neighbors, 
no tilt) agrees with experiments and with earlier theoretical predictions. 
Furthermore, we have discussed the transition to the fluid state, and concluded 
from the form of the pair correlation functions in the different phases, 
and from the way the transition temperature depends on the chain stiffness,
that the transition is mainly driven by the chains, again in agreement
with experimental\cite{barton} and theoretical\cite{rice,karaborni,me1} 
observations. 

\begin{figure}
\noindent
\fig{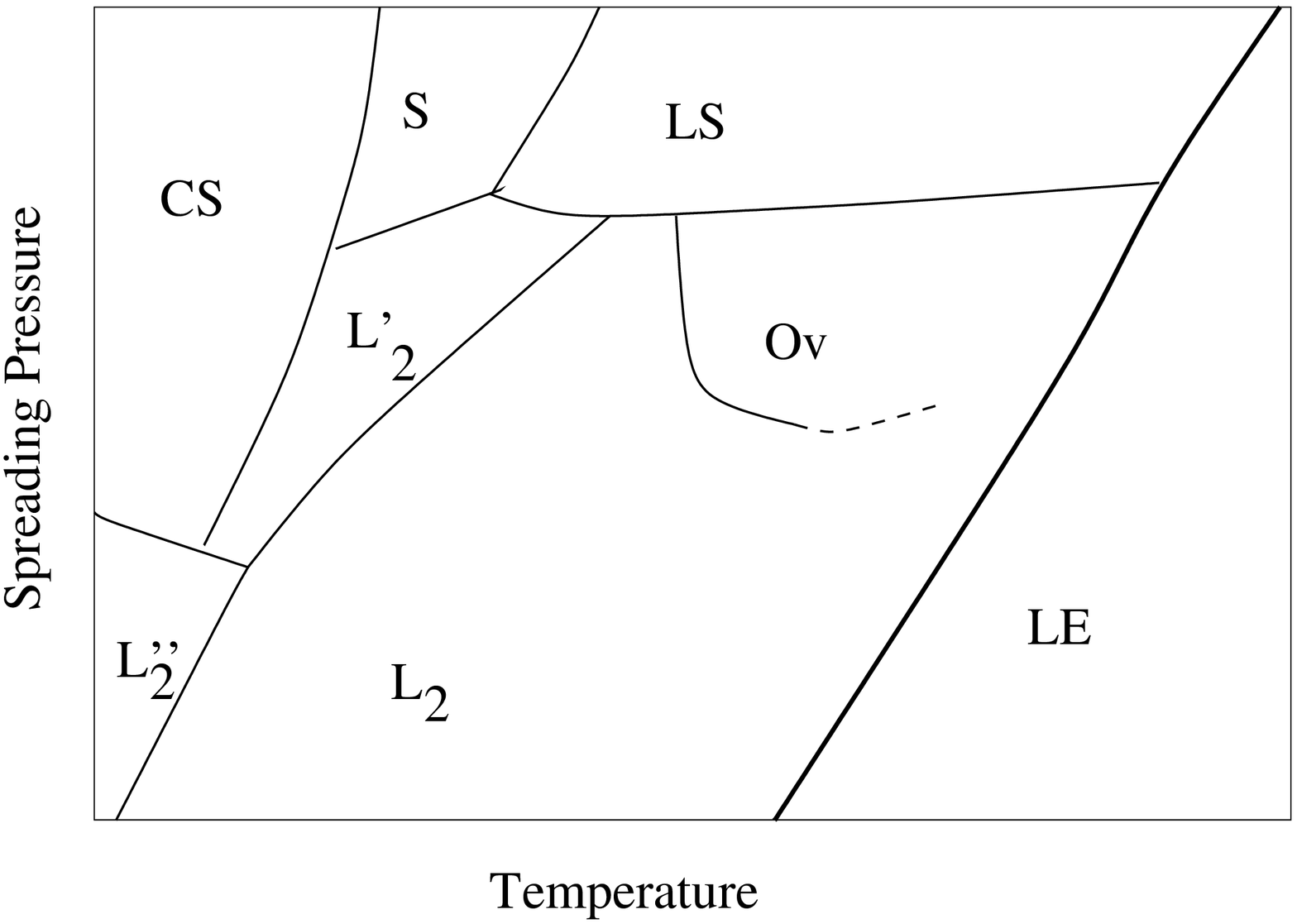}{100}{100} 
\caption{\label{fig1}}
\end{figure}

\begin{figure}
\noindent
\fig{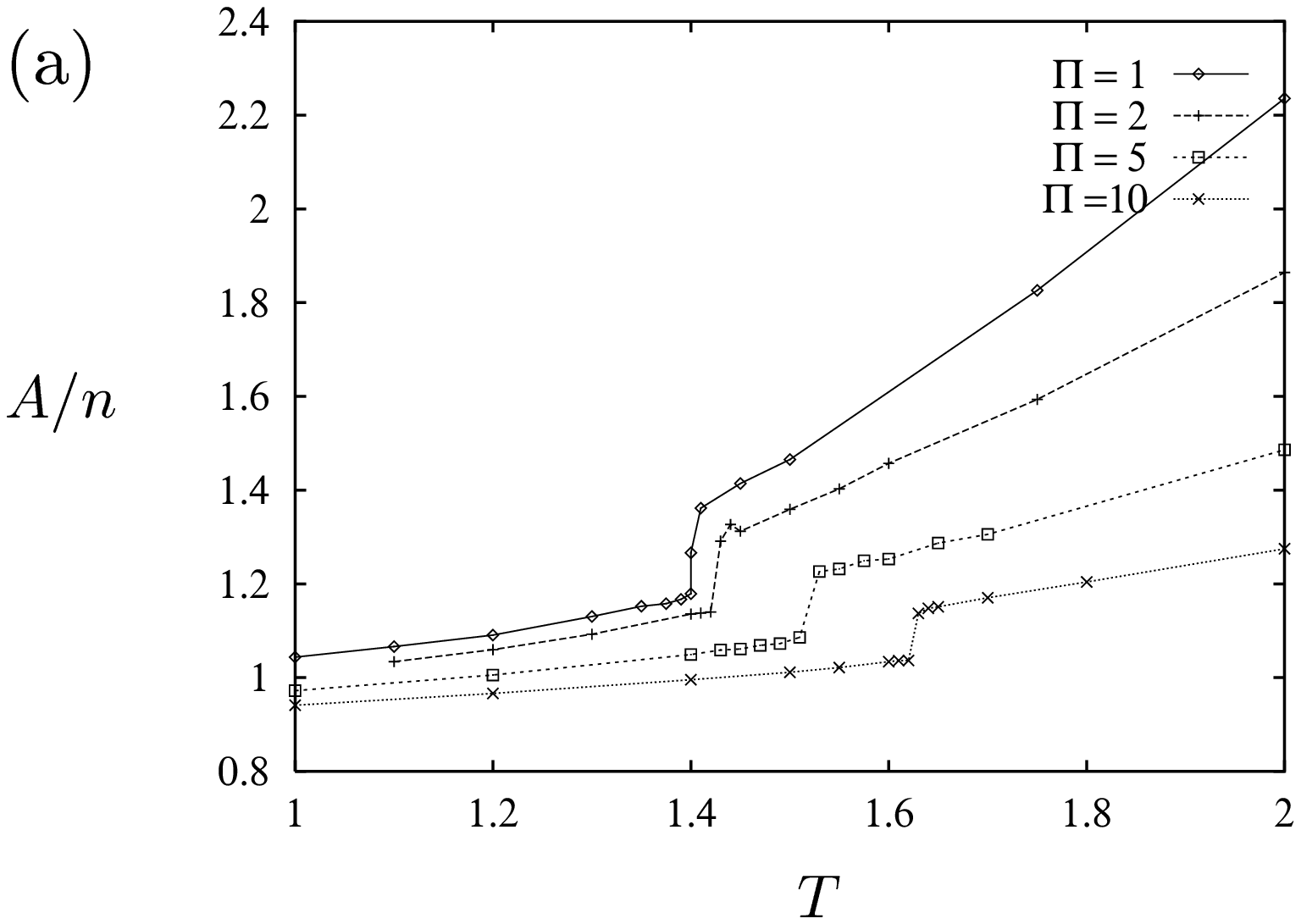}{150}{80} 
\fig{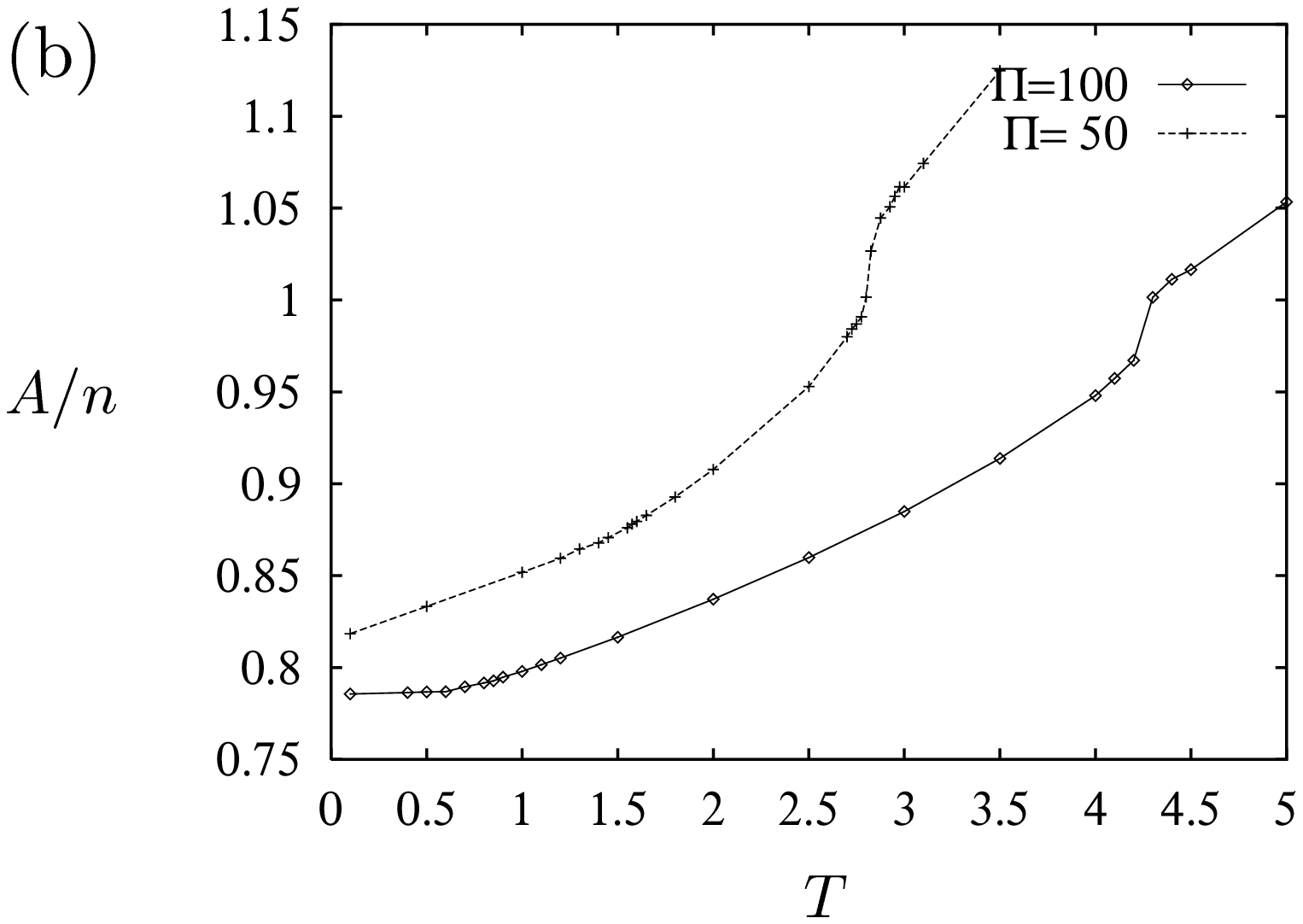}{150}{80}
\caption{\label{fig2}}
\end{figure}

\begin{figure}
\noindent
\fig{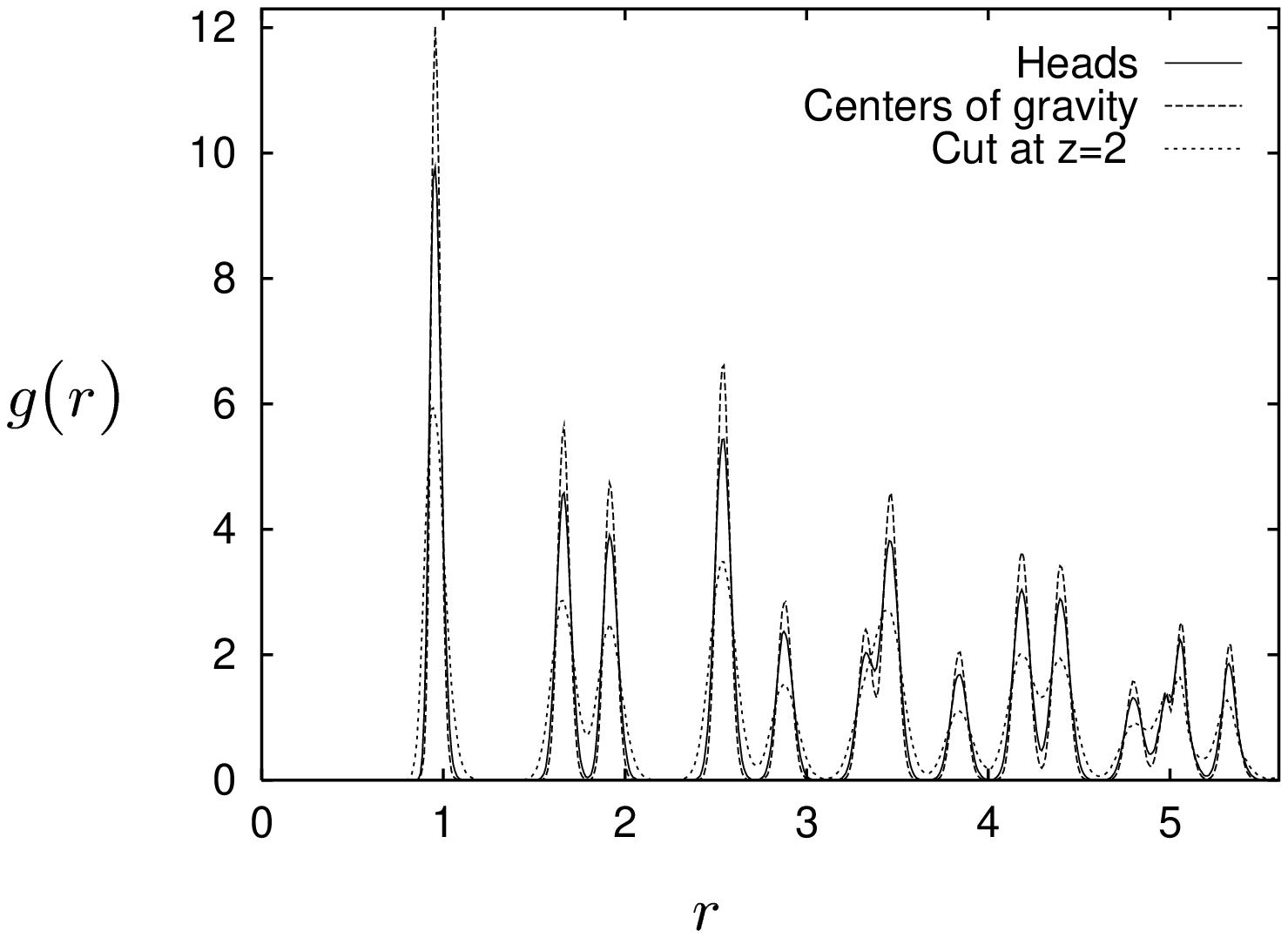}{150}{100}
\caption{\label{fig3}}
\end{figure}

\begin{figure}
\noindent
\fig{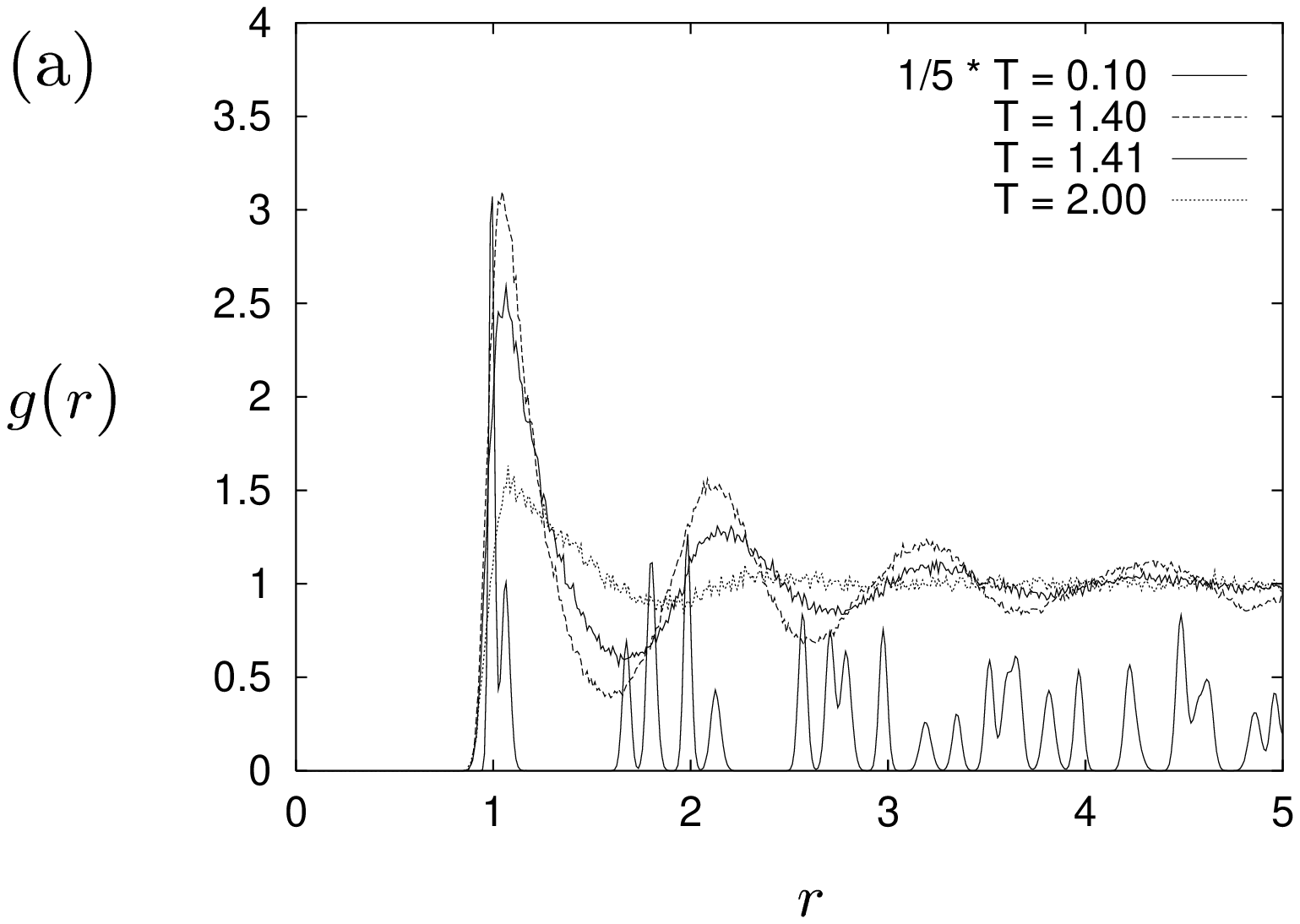}{150}{80}
\fig{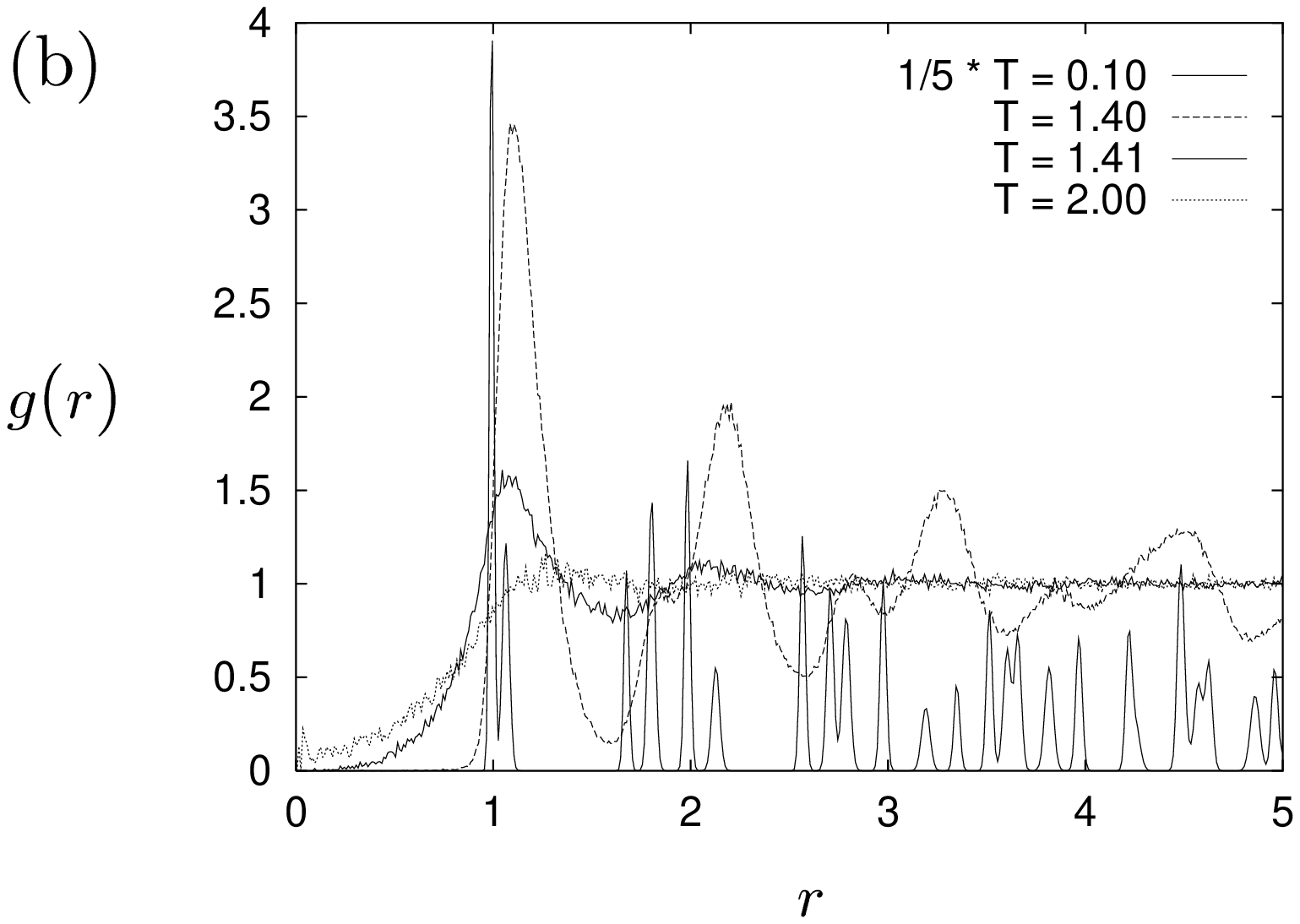}{150}{80}
\caption{\label{fig4}}
\end{figure}

\begin{figure}
\noindent
\fig{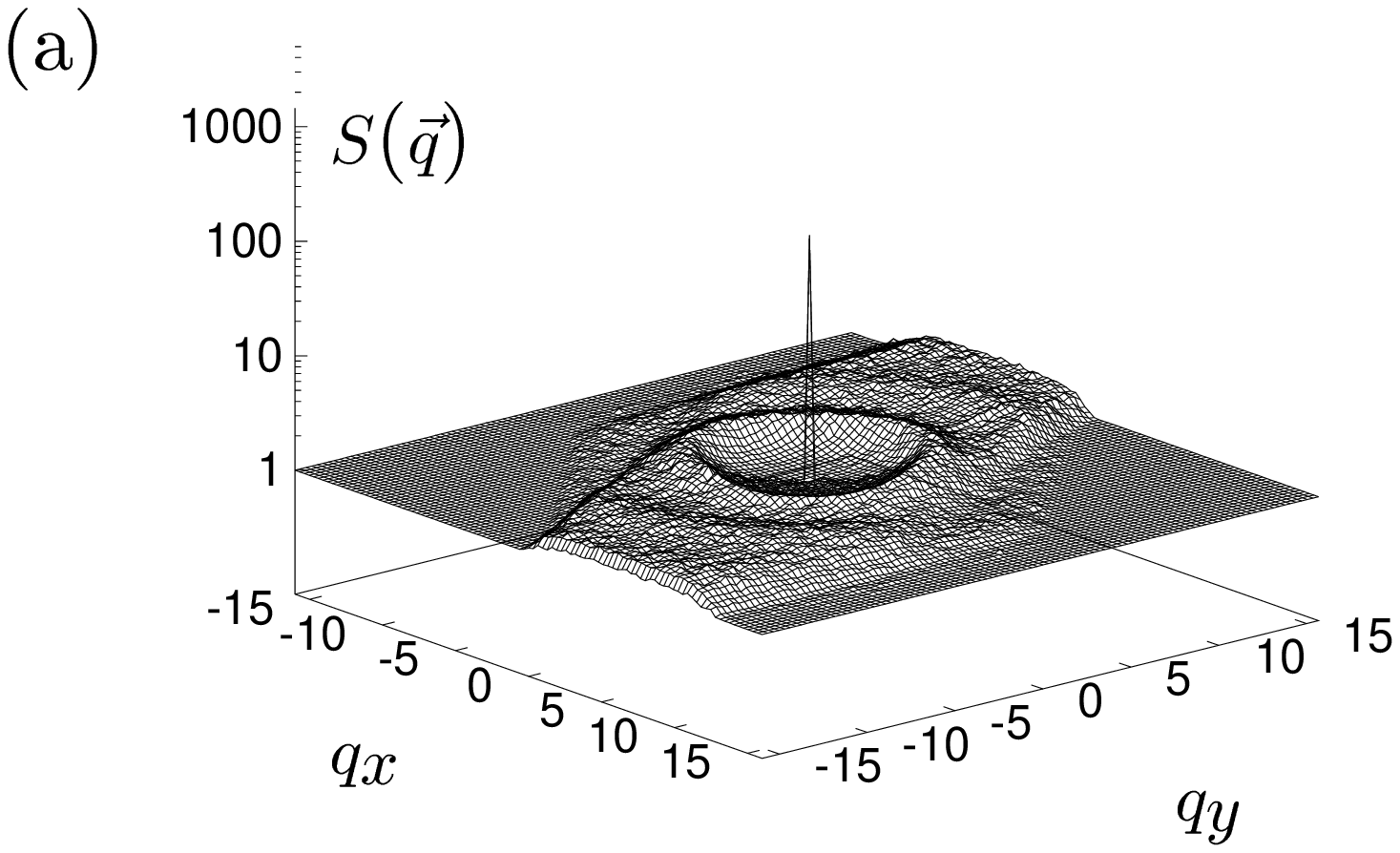}{150}{100}
\fig{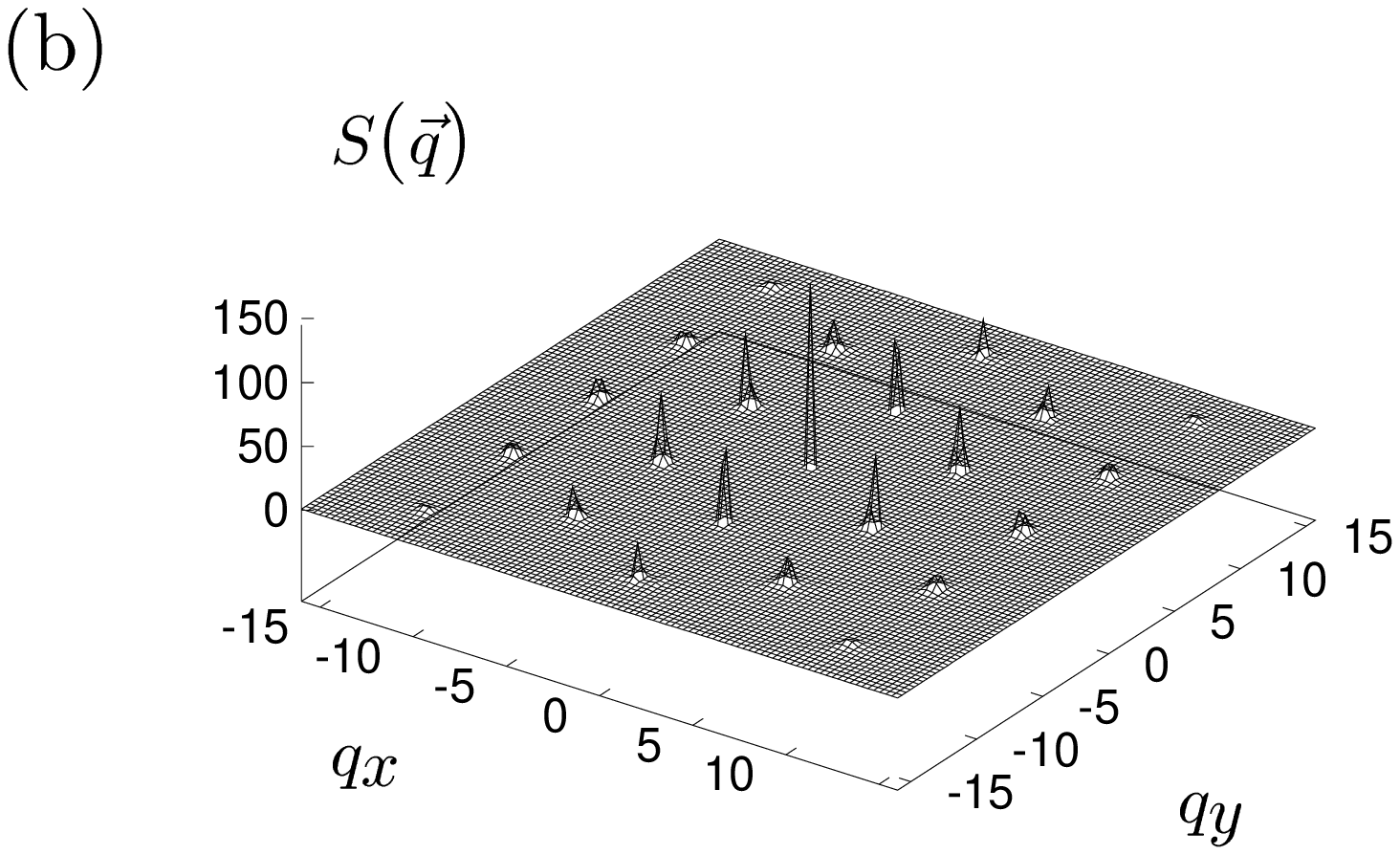}{150}{80}
\vspace{-1cm}
\caption{\label{fig5}}
\end{figure}

\begin{figure}
\noindent
\fig{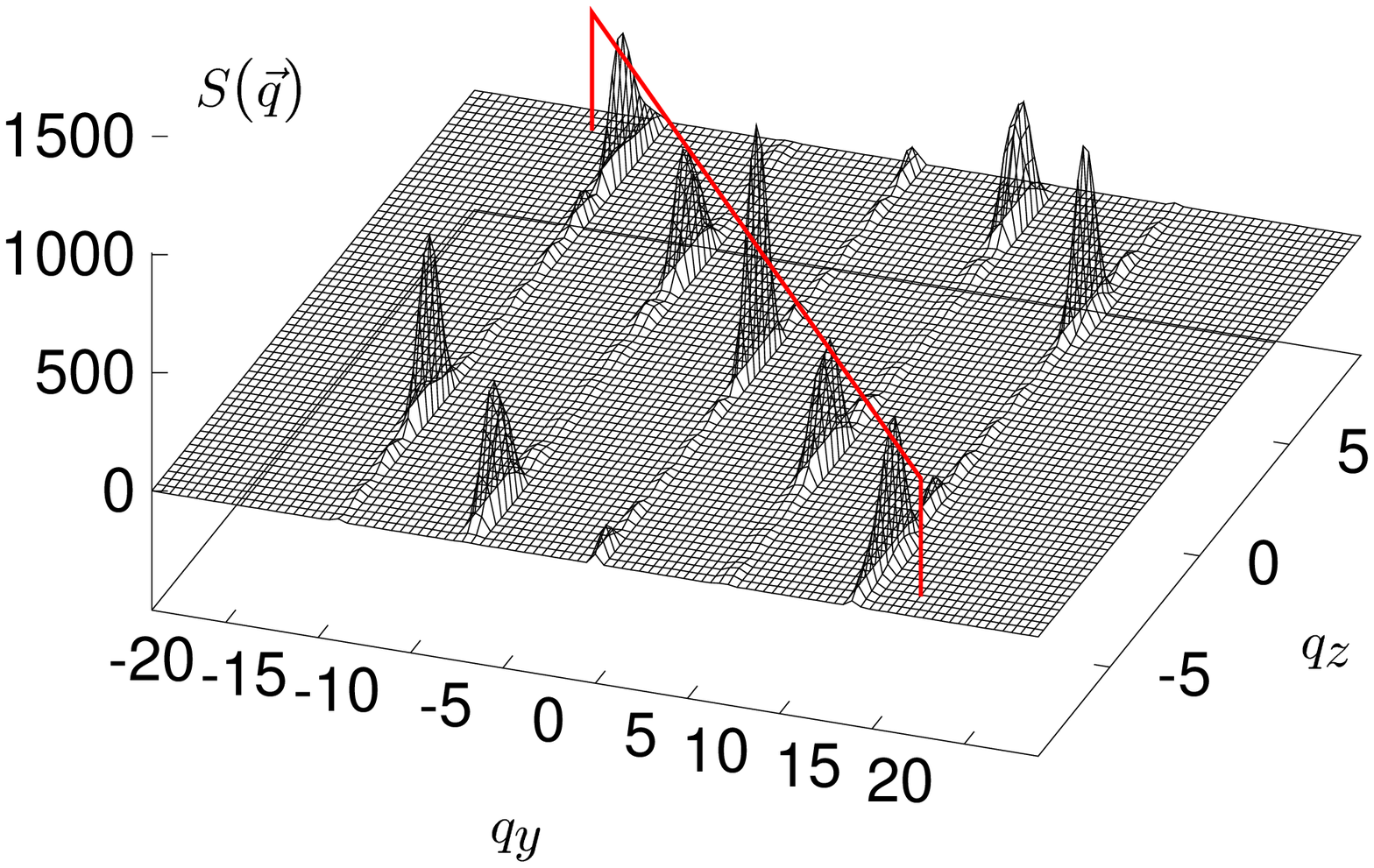}{150}{100}
\caption{\label{fig6}}
\end{figure}

\begin{figure}
\noindent
\fig{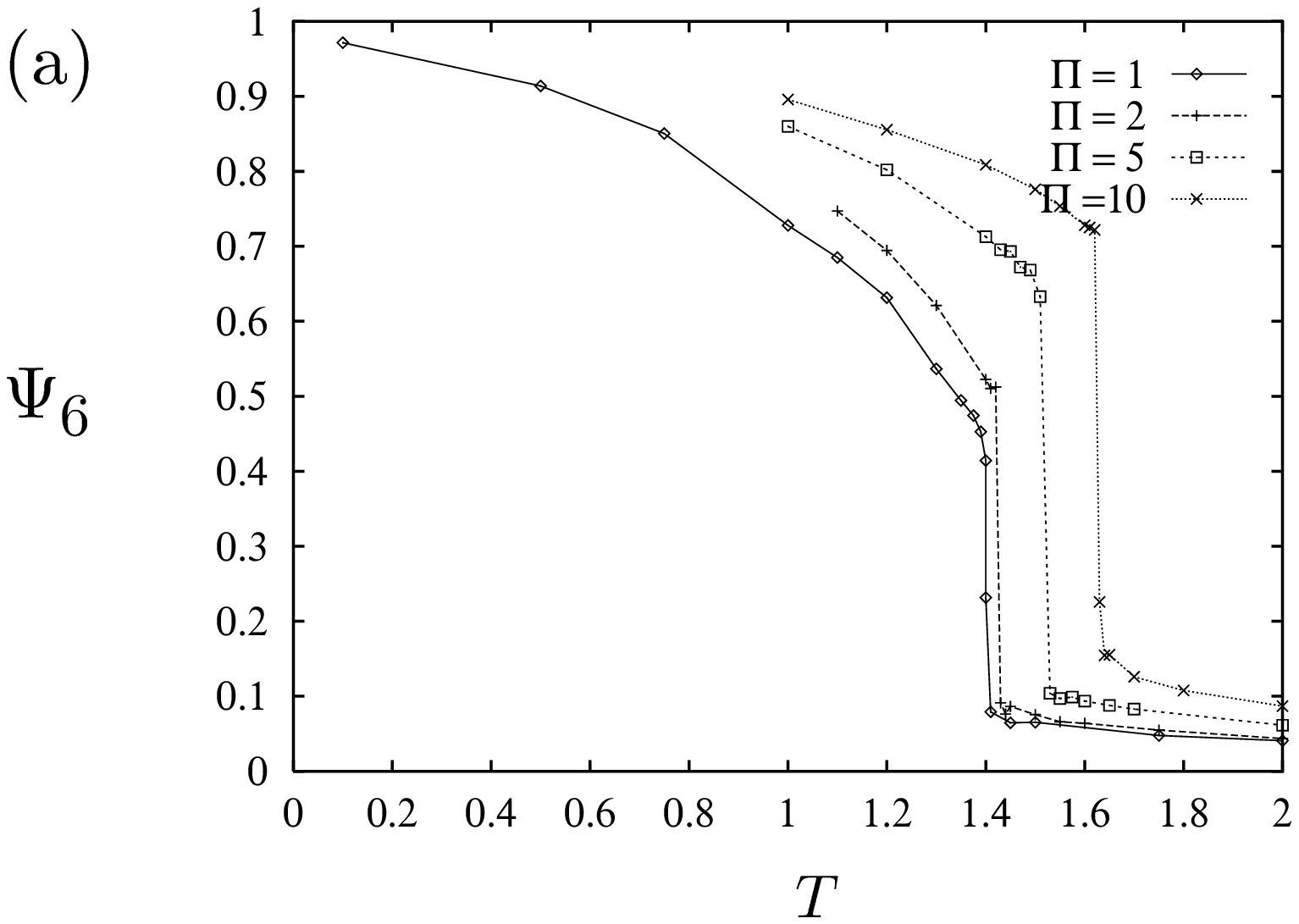}{150}{80}
\fig{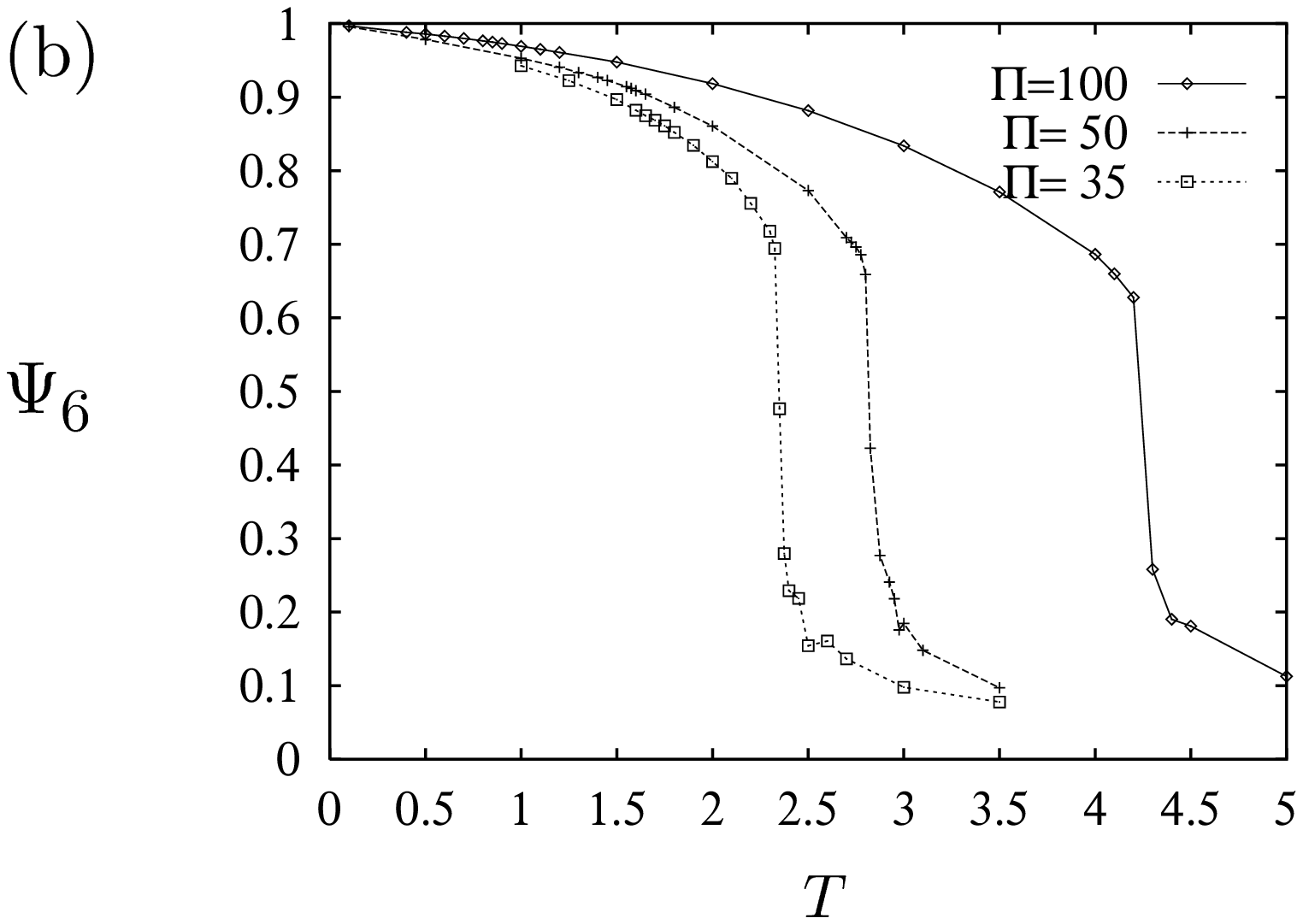}{150}{80}
\caption{\label{fig7}}
\end{figure}

\begin{figure}
\noindent
\fig{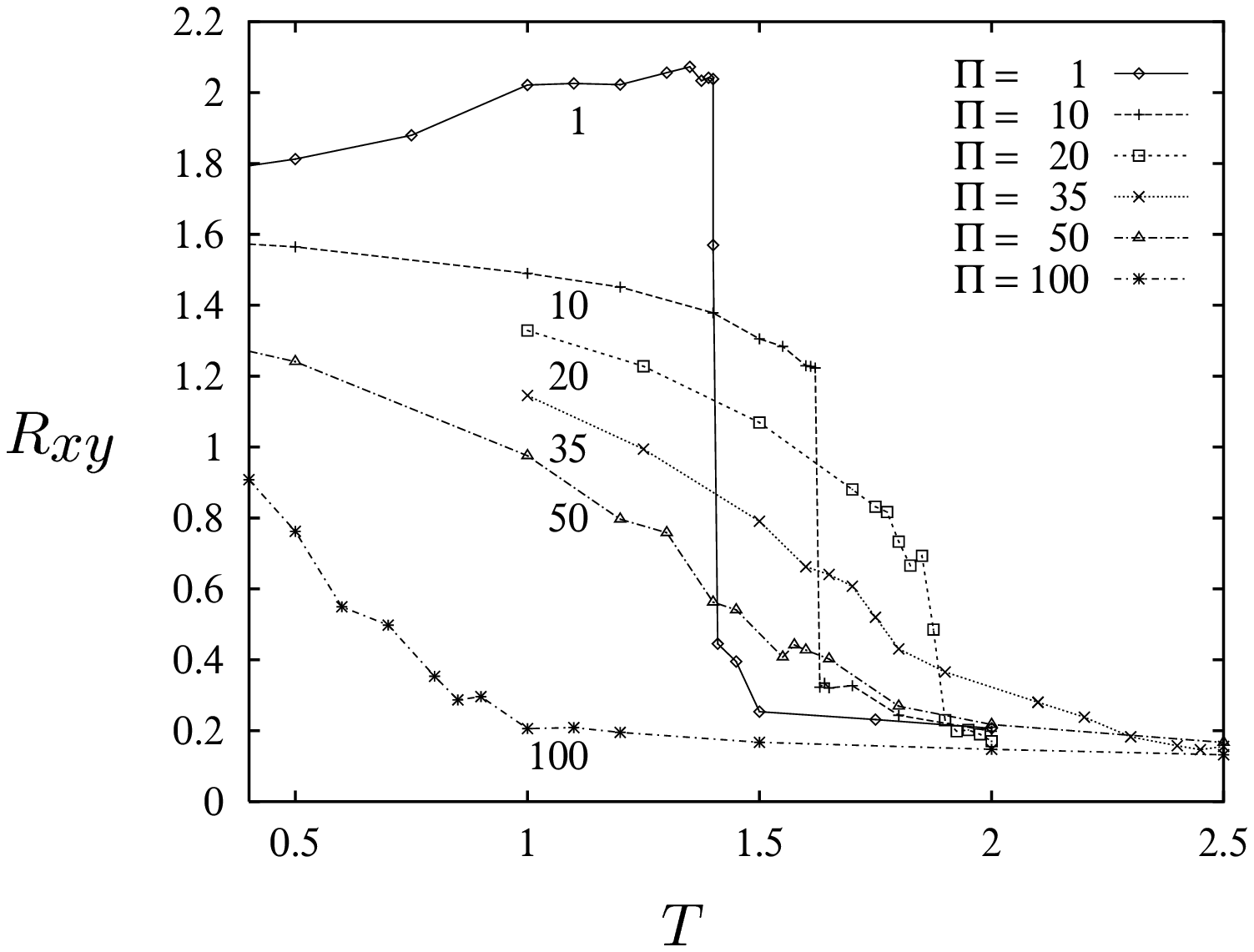}{150}{100}
\caption{\label{fig8}}
\end{figure}

\begin{figure}
\noindent
\fig{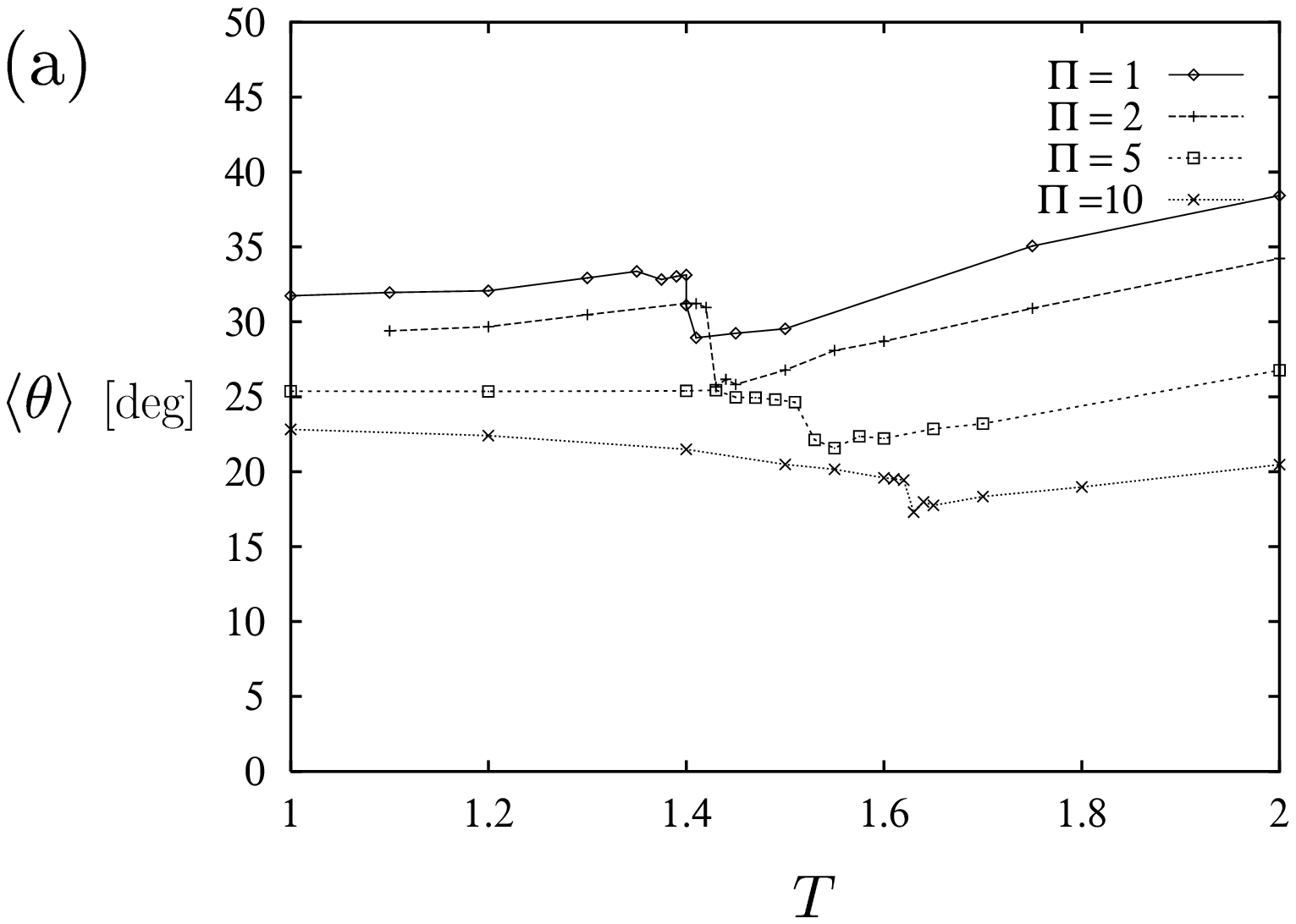}{150}{80}
\fig{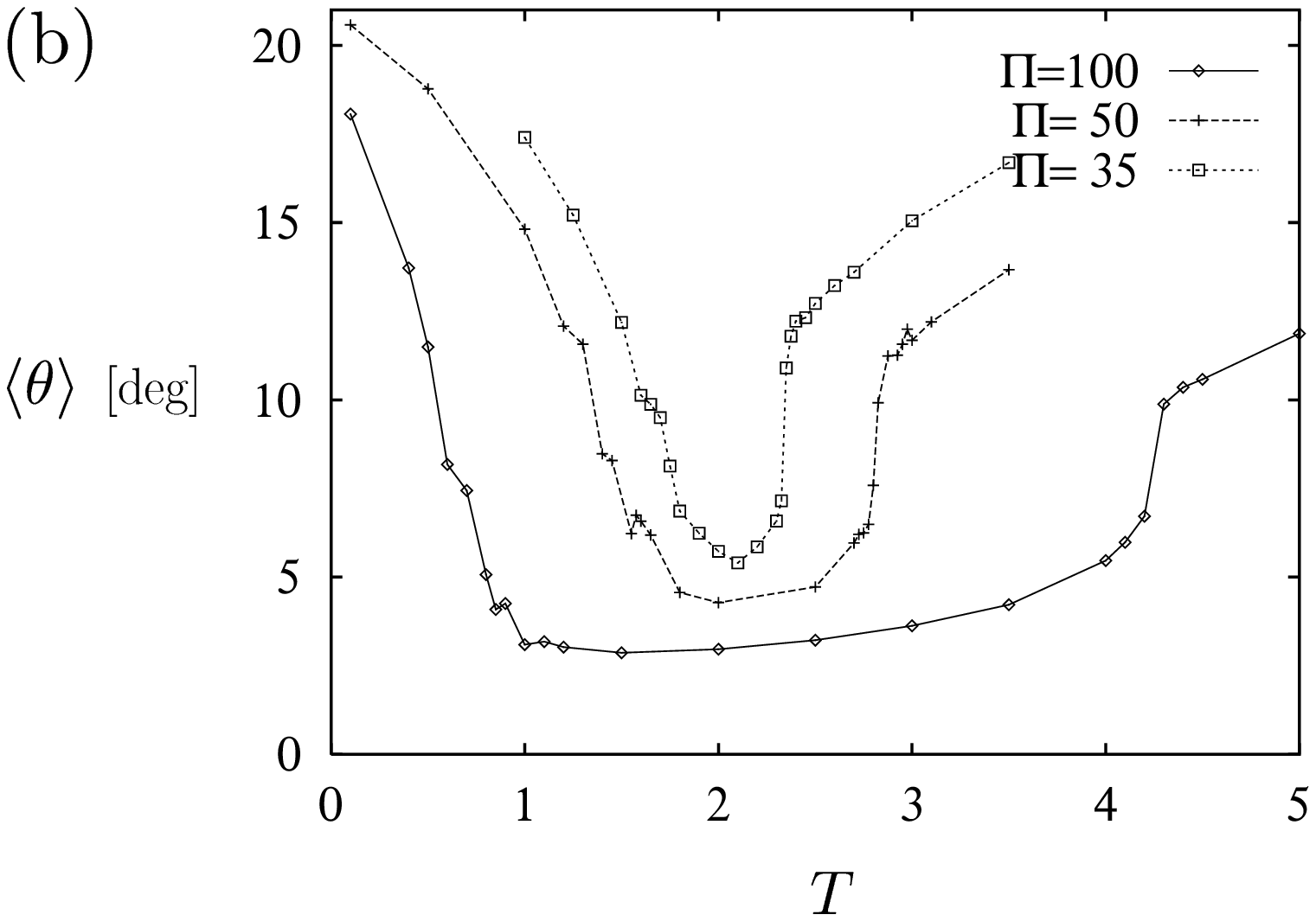}{150}{80}
\caption{\label{fig9}}
\end{figure}

\begin{figure}
\noindent
\fig{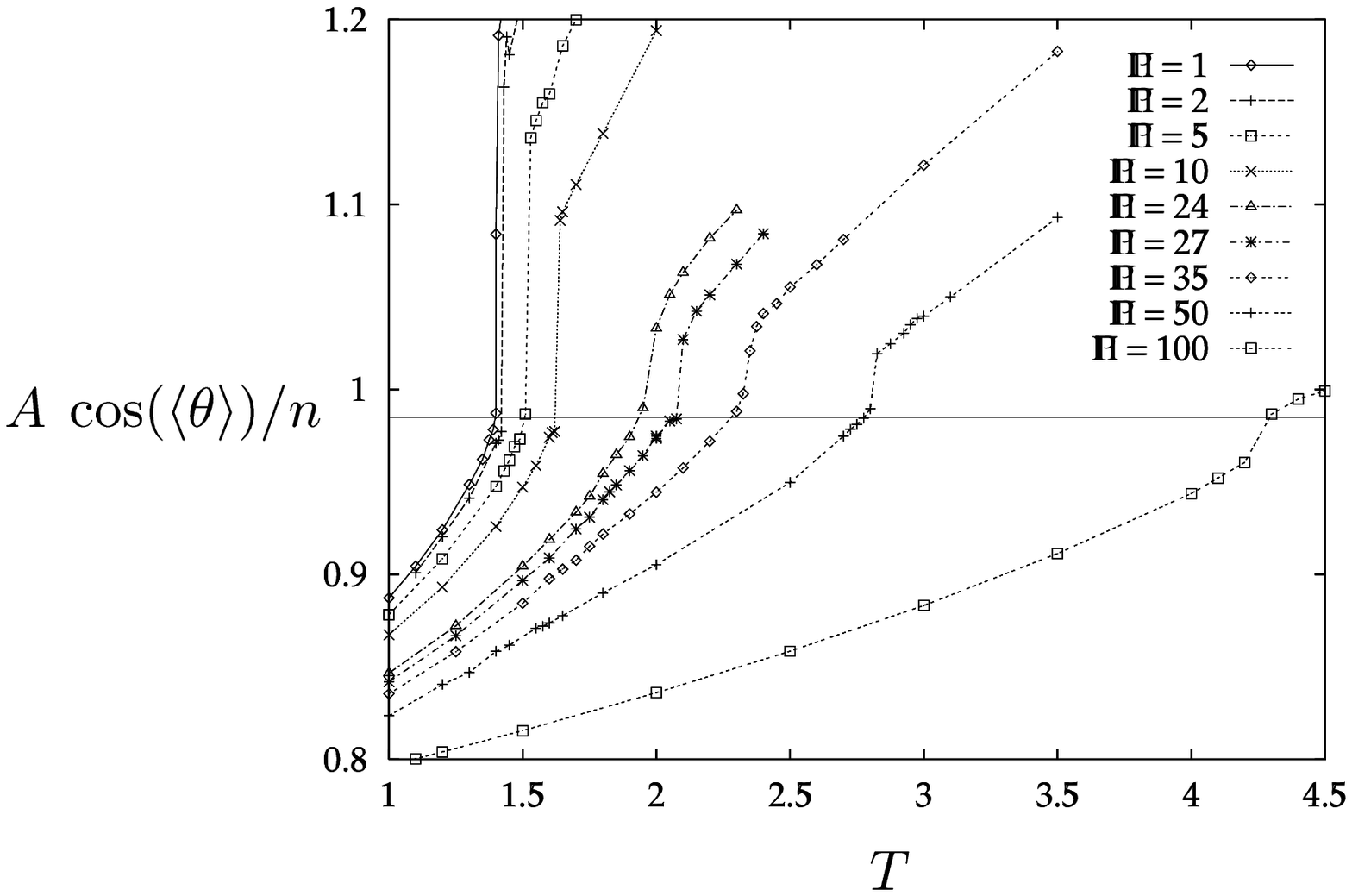}{150}{100}
\caption{\label{fig10}}
\end{figure}

\begin{figure}
\noindent
\fig{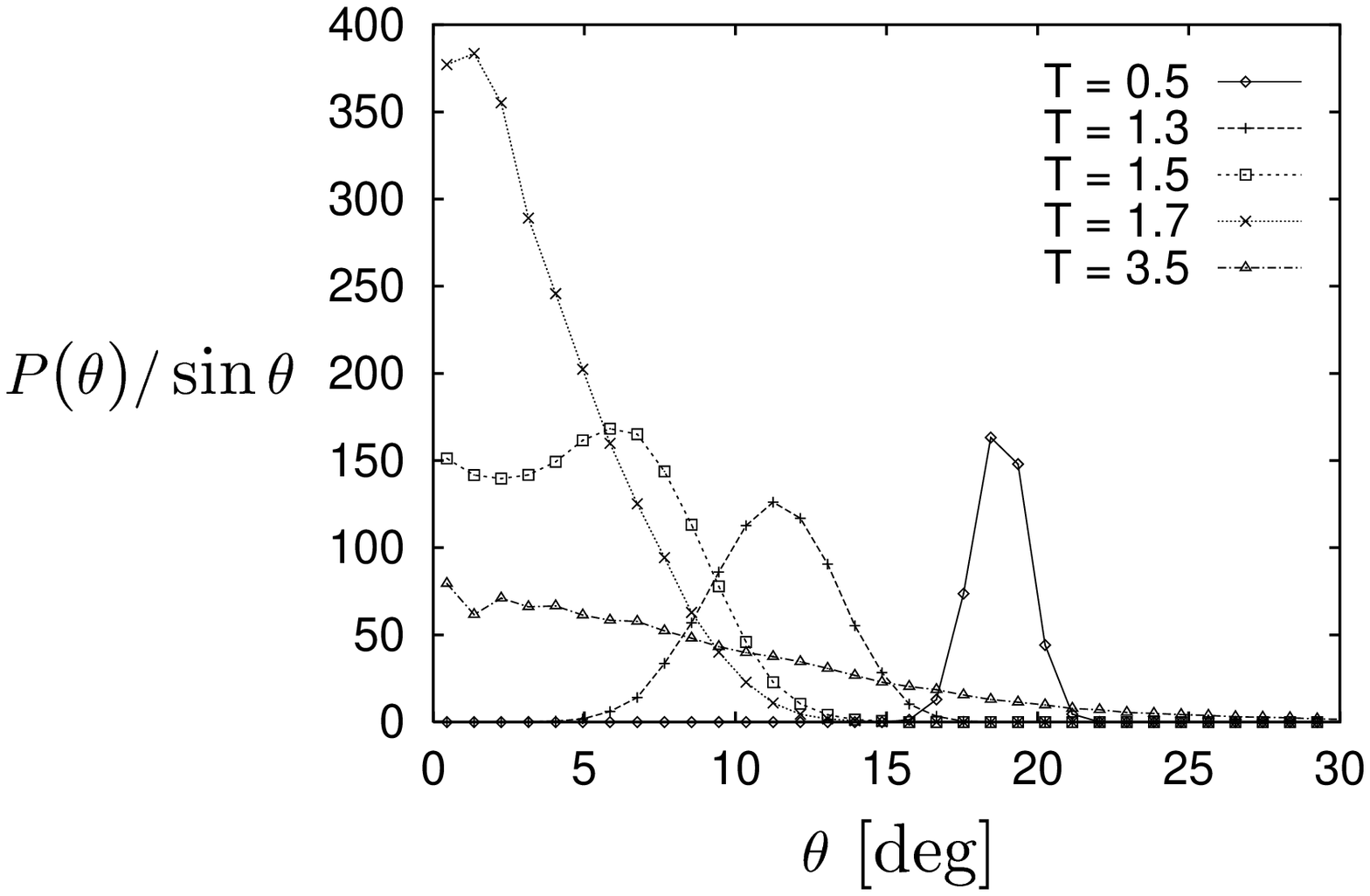}{150}{100}
\caption{\label{fig11}}
\end{figure}

\begin{figure}
\noindent
\fig{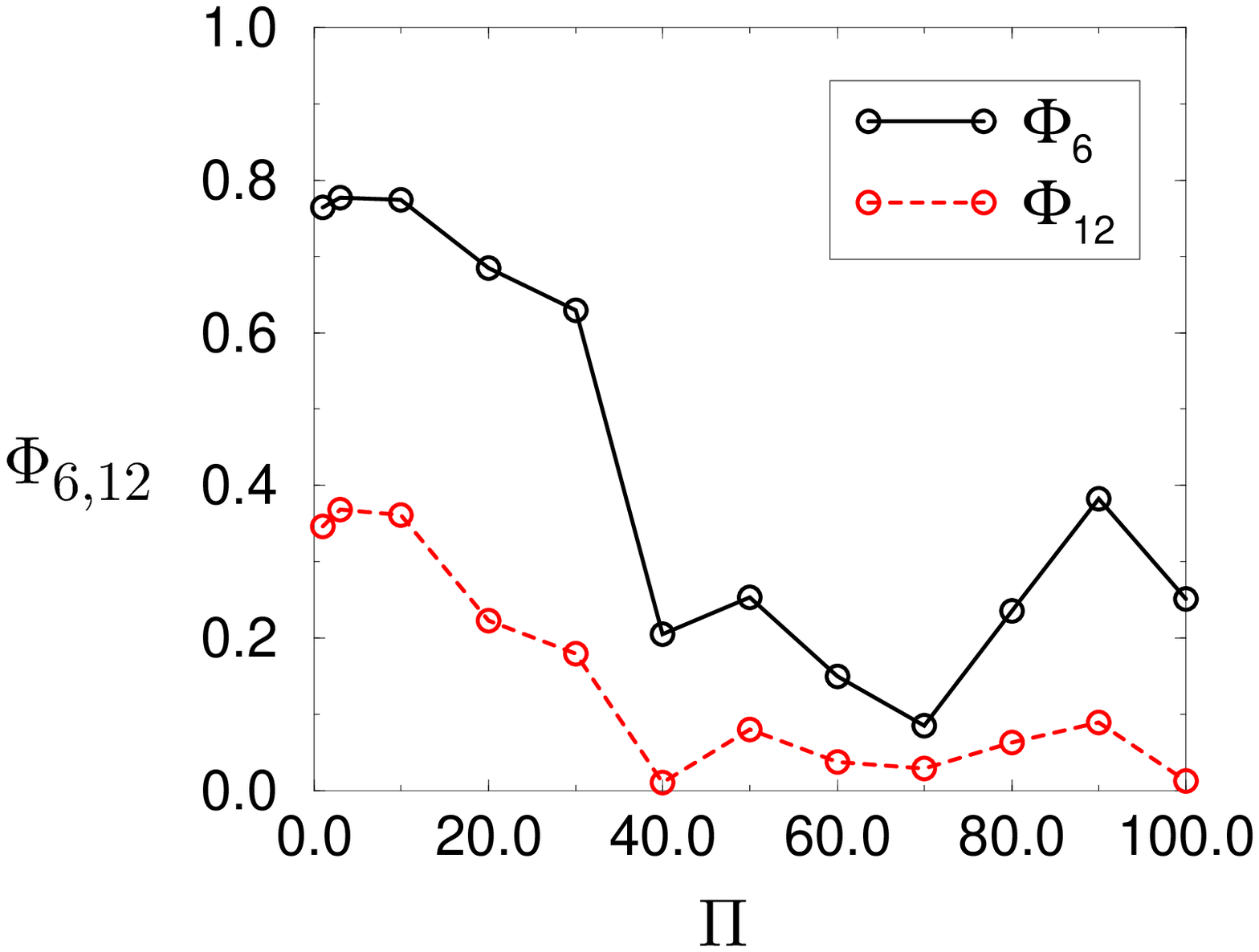}{150}{100}
\caption{\label{fig12}}
\end{figure}

\begin{figure}
\noindent
\fig{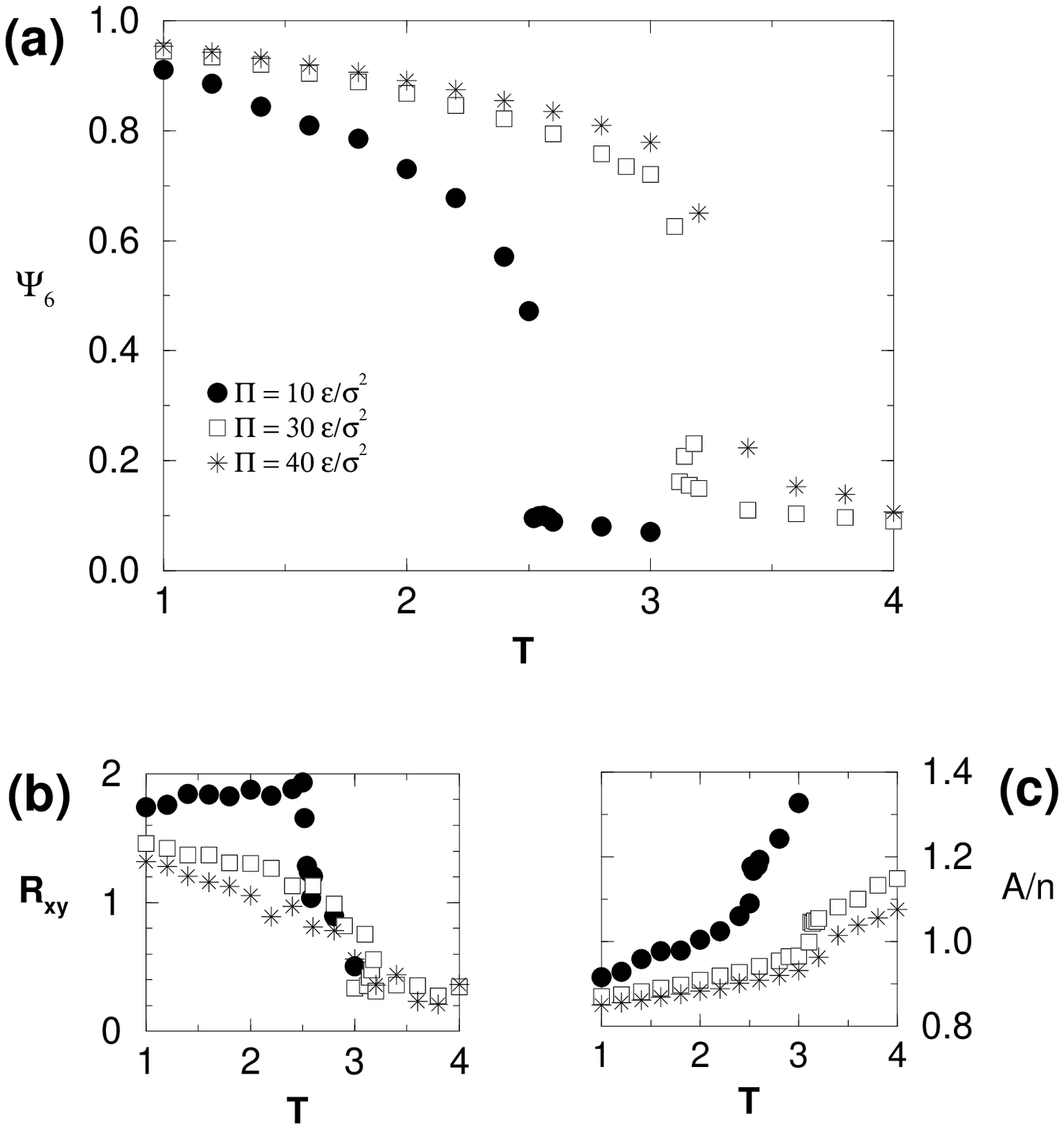}{180}{150} 
\caption{\label{fig13}}
\end{figure}

\begin{figure}
\noindent
\fig{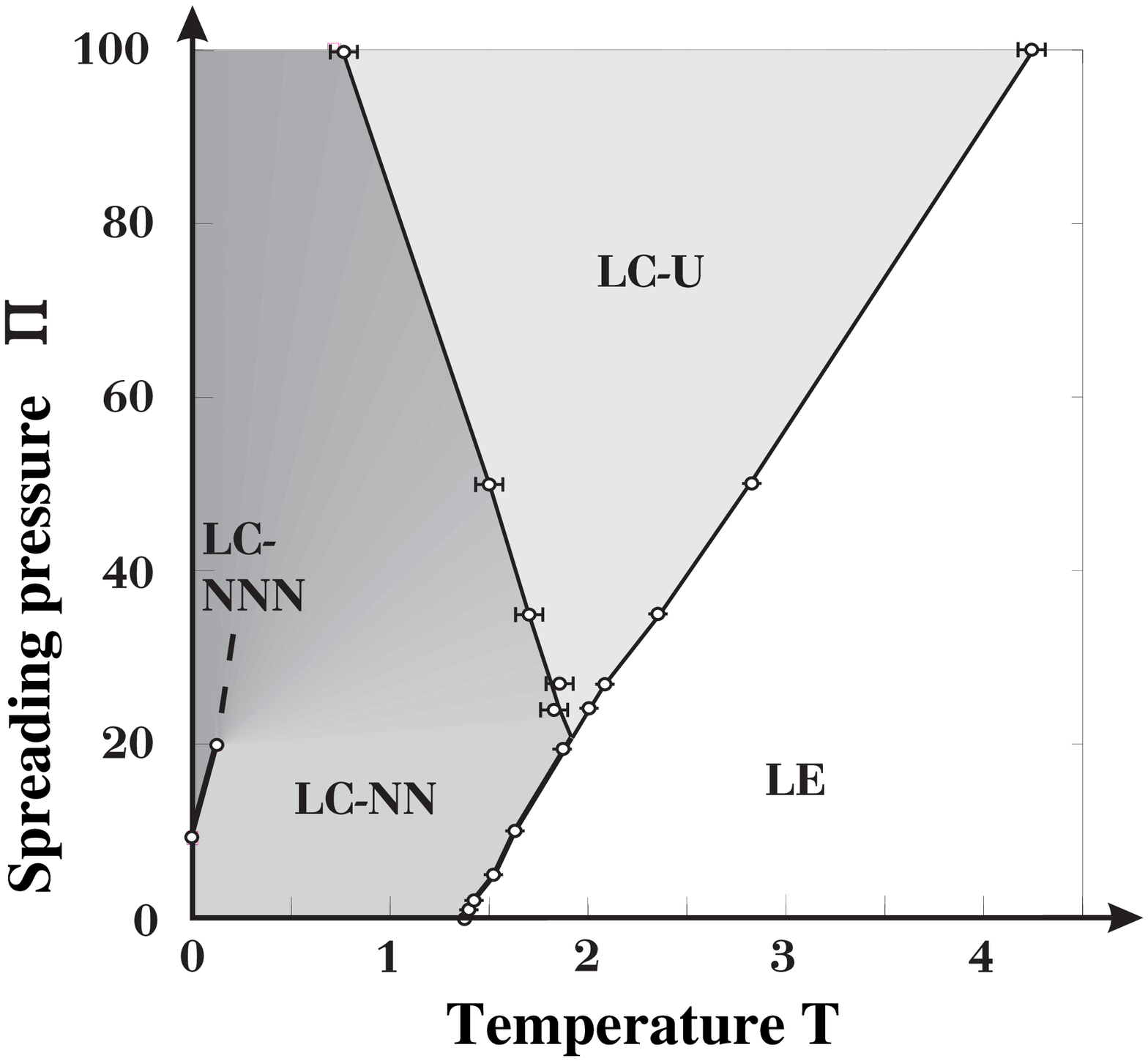}{150}{150}
\caption{\label{fig14}}
\end{figure}

\begin{figure}
\noindent
\fig{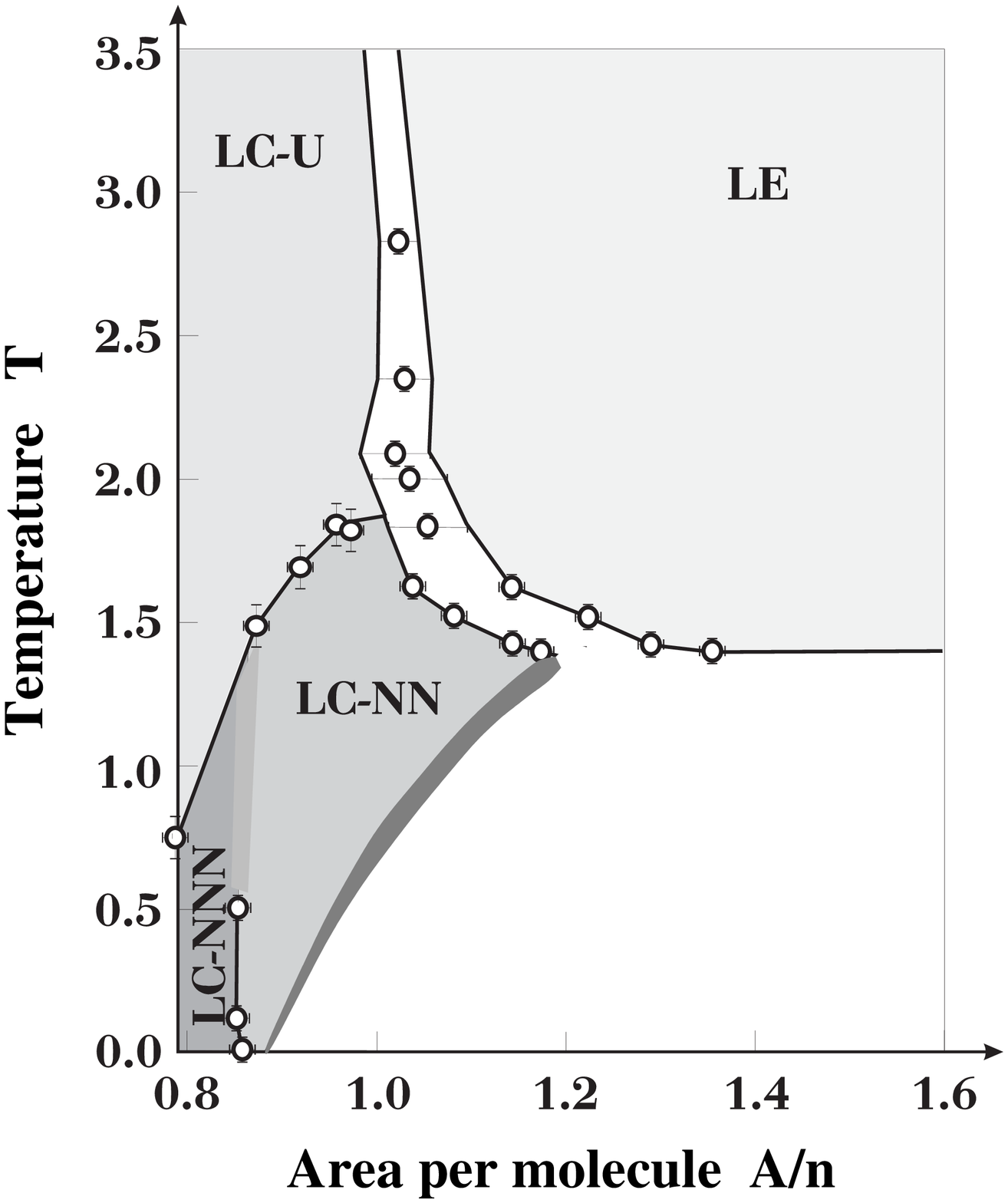}{150}{150} 
\caption{\label{fig15}}
\end{figure}

In other respect, the phase diagram is still quite different from the 
experimental one (Fig. \ref{fig1}). Some of the discrepancies are not 
surprising; for example, the model with its rotationally symmetric chains was 
never designed to reproduce the herringbone-ordered low-temperature structures.
Other differences are more interesting. The pressure at the transition from 
the tilted to the untilted phase decreases strongly with temperature, 
whereas it is almost independent of the temperature in experimental systems. 
Likewise, the transition pressure of the swiveling transition between
nearest neighbor tilt and next nearest neighbor tilt increases with
temperature, whereas the line separating the Ov and L${}_2$ phase in
Fig. \ref{fig1} moves to lower pressures.
This is presumably a consequence of the treatment of the head groups -- 
more specifically, of the rigid constraints which are imposed on them in the
model. The hard core interactions are much harder than the effective 
interactions between real head groups in water. Moreover, the heads in our 
model are confined to lie in a plane, whereas they can move in and out of 
the surface in real systems\cite{footnote2}. 

Further refinements of the model will thus have to focus on the representation 
of the head groups. We have already mentioned the interplay between head size, 
spreading pressure, and tilting transitions. A more detailed study of the 
influence of the head size on the phase behavior shall be presented 
elsewhere\cite{christoph2}. Future work will be concerned with the
effect of relaxing some of the constraints on the head groups, {\em i.e.},
giving them additional degrees of freedom in the $z$-direction, and
possibly softening the interactions between them. One could also think
of introducing interactions between the tails and the substrate. However,
the tails hardly come into contact with the substrate at most densities of
interest, therefore this will probably not change the phase behavior
significantly.

On the other hand, we have seen that already the present simple model 
reproduces many important properties of amphiphile monolayers.
Hence it can be used as a starting point for further investigations.
In particular, simulations of much larger systems and a systematic
variation of system sizes would be desirable to shed light on some of the
questions which have remained open in the present study. These would
help to elucidate the exact nature of the tilting transitions and the
order of the melting transition, to examine the unlocked tilted state,
and to clarify whether our model actually does exhibit hexatic phases.

\section*{Acknowledgments}

We are grateful to Kurt Binder for numerous enlightening discussions,
and to Frank Martin Haas and Rudolf Hilfer for sharing with us
their expertise on the simulation of these systems, and in particular for 
letting us have their simulation code, which has been an invaluable
starting point for the development of our own. We have benefitted from useful 
interactions with Burkhard D\"unweg, Peter Nielaba, and Nigel Wilding. 
F.S. is supported by the Deutsche Forschungsgemeinschaft through a
Heisenberg fellowship, and C.S. through a Ph.D. studentship associated with
the Graduiertenkolleg on supramolecular systems in Mainz.

\clearpage

\noindent
Generic phase diagram for fatty acid monolayers. The phases 
$LS$, $S$ and $CS$ are on average untilted, whereas $Ov$ and $L_2'$
show tilt towards next nearest neighbors, and $L_2$, $L_2''$ towards
nearest neighbors. In $CS$, $S$, $L_2'$ and $L_2''$, the backbones
of the hydrocarbon chains are ordered. In $CS$ and $L_2''$, the
molecules have crystalline order in addition.

\clearpage

\noindent
Area per molecule $A/n$ in units of  $\sigma^2$ vs. temperature $T$ in
units of $\epsilon/k_B$ for a choice of low (a) and high (b) 
pressures $\Pi$ (in units of $\epsilon/\sigma^2$) as indicated.

\clearpage

\noindent
Radial pair correlation functions $g(r)$ vs. $r$ in units of $\sigma$
at pressure $\Pi =100 \epsilon/\sigma^2$ and temperature $T=1 \epsilon/k_B$.
Correlation functions are shown for the heads (solid line), for the points 
where the molecules cross the plane at $z=2 \sigma$ above the surface 
(dotted line), and for the projection of the center of gravity onto the 
$xy$ plane (dashed line). The values of $g(r)$ for $T=0.1 \epsilon/k_B$ are 
divided by a factor of five for the clarity of presentation.

\clearpage

\noindent
Radial pair correlation functions $g(r)$ vs. $r$ in units of $\sigma$
at pressure $\Pi=1 \epsilon/\sigma^2$ and various temperatures as
indicated. Correlation functions are shown for the heads (a),
and for projections into the $xy$ plane of the centers of gravity (b).
Temperatures $T$ are given in units of $\epsilon$.
The correlation functions $g(r)$ for the temperature $T=0.1 \epsilon$ have 
been divided by a factor of 5.

\clearpage

\noindent
Structure factor $S(\vec{q})$ in the $xy$ plane ($q_z=0$)
for a disordered state (a)
and an untilted ordered state (b). 
Parameters are $\Pi=10 \epsilon/\sigma^2$,
$T=2.5 \epsilon/k_B$ in (a), and $\Pi = 50 \epsilon/\sigma^2$,
$T=2.0 \epsilon/k_B$ in (b).

\noindent
Structure factor $S(\vec{q})$ in the $yz$ plane ($q_x=0$)
for an ordered state with
tilt towards next nearest neighbors. Parameters are $\Pi=50 \epsilon/\sigma^2$ 
and $T=0.1 \epsilon/k_B$.

\clearpage

\noindent
Order parameter $\Psi_6$ vs. temperature $T$ in units of $\epsilon/k_B$
for different pressures $\Pi$ (in units of $\epsilon/\sigma^2$) as indicated.

\clearpage

\noindent
Order parameter $R_{xy}$ in units of $\sigma^2$ vs. temperature $T$ in units 
of $\epsilon/k_B$ 
for different pressures $\Pi$ (in units of $\epsilon/\sigma^2$) as indicated.

\clearpage

\noindent
Average tilt angle $\langle \theta \rangle $ 
in degrees vs. temperature $T$ in units of 
$\epsilon/k_B$ for different pressures $\Pi$ (in units of $\epsilon/\sigma^2$) 
as indicated.

\noindent
Product of the cosine of the average tilt angle with the area per molecule,
$A \cos(\langle \theta \rangle)/n$, in units of $\sigma^2$, 
vs. temperature $T$ in units of $\epsilon/k_B$ for different pressures $
\Pi$ (in units of $\epsilon/\sigma^2$) as indicated.
The horizontal line indicates the position of $a_c=0.985 \sigma^2$.

\noindent
Histogram $P(\theta)/\sin \theta $ of the tilt angle $\theta$ (in degrees)
at pressure $\Pi = 50 \epsilon/\sigma^2$ for different temperatures 
(in units of $\epsilon/k_B$).

\clearpage

Order parameters $\Phi_6$ and $\Phi_{12}$ vs. pressure $\Pi$ in units of 
$\epsilon/\sigma^2$ at temperature $T=0.5 \epsilon/k_B$.

\clearpage

Order parameters $\Psi_6$ and $R_{xy}$, and area per molecule $A/n$ 
vs. temperature $T$ in units of $\epsilon/k_B$ for pressures 
$\Pi=10 \epsilon/\sigma^2$ (filled circles), 
$30 \epsilon/\sigma^2$ (open squares),
$40 \epsilon/\sigma^2$ (stars) in systems of stiff chains, 
($k_A=100 \epsilon$).

Phase diagram in the pressure-temperature plane. Pressure $\Pi$ is given
in units of $\epsilon/\sigma^2$, and temperature $T$ in units of $\epsilon/k_B$.
LE denotes disordered phase, LC-NN ordered phase with tilt towards nearest 
neighbors, LC-NNN ordered phase with tilt towards next nearest neighbors, 
and LC-U untilted ordered phase. The transition between LC-NN and LC-NNN 
could not be located at pressures above $\Pi = 20 \epsilon/\sigma^2$. 
See text for more explanation.

Phase diagram in the area-temperature plane. Area per molecule $A/n$ is given
in units of $\sigma^2$, and temperature $T$ in units of $\epsilon/k_B$.
LE denotes disordered phase, LC-NN ordered phase with tilt towards nearest 
neighbors, LC-NNN ordered phase with tilt towards next nearest neighbors, 
and LC-U untilted ordered phase. See text for more explanation.


\begin{thebibliography}{99}
\lettersize
\bibitem{mono-reviews} 
  G. M. Bell, L. L. Combs, L. J. Dunne, Chem. Rev. {\bf 81}, 15 (1981);
  H. M\"ohwald, Ann. Rev. Phys. Chem. {\bf 41}, 441 (1990);
  H. M. McConnell, ibid {\bf 42}, 171 (1991);
  C. M. Knobler, R.C. Desai, ibid {\bf 43}, 207 (1992).
  D. Andelman, F. Brochard, C.M. Knobler, F. Rondelez,
  in {\em Micelles, Membranes, Microemulsions and Monolayers}, p. 559,
  W. M. Gelbart, D. Roux and A. Ben-Shaul eds. (Springer, 994).
\bibitem{lb-books}
  W. A. Barlow, {\em Langmuir-Blodgett Films} (Elsevier, 1980);
  M. C. Petty, {\em Langmuir-Blodgett Films} (Cambridge University Press, 1996).
\bibitem{kaganer-review}
  V. M. Kaganer, H. M\"ohwald, P. Dutta, Rev. Mod. Phys., in press (1998).
\bibitem{selinger}
  J.V. Selinger, Z. Wang, R.F. Bruinsma, C.M. Knobler,
  Phys. Rev. Lett. {\bf 70}, 1139 (1993);
\bibitem{kaganer}
  V. M. Kaganer, E. B. Loginov, 
    Phys. Rev. Lett. {\bf 71}, 2599 (1993); Phys. Rev. E {\bf 51}, 2237 (1995);
  V. M. Kaganer, V. L. Indenbom, J. Phys. II France {\bf 3}, 813 (1993).
\bibitem{selinger2}
  J. V. Selinger, D. R. Nelson, Phys. Rev. A {\bf 39}, 3135 (1989).
\bibitem{kox}
 A. J. Kox, Nature {\bf 285}, 317 (1980);
 S. H. Northrup, M. S. Curvin, J. Phys. Chem. {\bf 89}, 4707 (1985).
 M. A. Moller, D. J. Tildesley, K. S. Kim, N. Quirke, 
    J. Chem. Phys. {\bf 94} 8390 (1991).
 P. Ahlstr\"om, J. C. Berendsen, J. Phys. Chem. {\em 97} 13691 (1992).
\bibitem{klein}
 J. P. Bareman, G. Cardini, M. L. Klein. Phys. Rev. Lett. {\bf 60}, 2152 (1988).
 G. Cardini, J. P. Bareman, M. L. Klein. Chem. Phys. Lett. {\bf 145}, 493 (1988);
 J. P. Bareman, M. L. Klein. J. Phys. Chem. {\bf 94}, 5202 (1990).
 J. Hautman, M. L. Klein, 
   J. Chem. Phys. {\bf 91}, 4994 (1989); {\bf 93}, 7483 (1990).
\bibitem{alper}
 H. E. Alper, D. Bassolino, T. R. Stouch. 
    J. Chem. Phys. {\bf 98}, 9798 (1993); {\bf 99}, 5547 (1993).
\bibitem{rice}
 J. Harris, S. A. Rice, J. Chem. Phys. {\em 89}, 5898 (1988).
 S. Shin, N. Collazo, S.A. Rice,
     J. Chem. Phys. {\bf 96}, 1352 (1991); {\bf 98}, 3469 (1992);
 N. Collazo, S. Shin, S.A. Rice, J. Chem. Phys. {\bf 96}, 4735 (1991);
 J. Gao, S. A. Rice, J. Chem. Phys. {\bf 99}, 7020 (1993);
 M. E. Schmidt, S. Shin, S. A.  Rice, 
   J. Chem. phys. {\bf 104} 2101 and 2114 (1996);
 K.-P. Bell, S. A. Rice. 
   J. Chem. Phys. {\bf 99} 4160 (1993); {\bf 104}, 1684 (1996).
\bibitem{karaborni}
 S. Karaborni, S. Toxvaerd, 
   J. Chem. Phys. {\bf 96}, 5505 (1992); {\bf 97}, 5876 (1992);
 S. Karaborni, S. Toxvaerd, O. H. Olsen, J. Phys. Chem. {\bf 96}, 4965 (1992);
 S. Karaborni, Langmuir {\bf 9}, 1334 (1993);
 S. Karaborni, G. Verbist, Europhys. Lett. {\bf 27}, 467 (1996);
 J. I. Siepmann, S. Karaborni, M. L. Klein, J. Phys. Chem. {\bf 98}, 6675 (1994).
\bibitem{binder-tilt}
 M. Kreer, K. Kremer, K. Binder, J. Chem. Phys. {\bf 92}, 6195 (1990);
 M. Scheringer, R. Hilfer, K. Binder, J. Chem. Phys. {\bf 96}, 2269 (1991);
\bibitem{tilt} 
 Z. Cai, S.A. Rice,
   Faraday Discuss. Chem. Soc. {\bf 89}, 211 (1990);
   J. Chem. Phys. {\bf 96}, 6229 (1992);
 V. M. Kaganer, M. A. Opisov, I. R. Peterson,
   J. Chem. Phys. {\bf 98}, 3512 (1992);
 B. Pal, S. Modak, A. Datta, Surf. Science {\bf 310}, 407 (1993);
 S. M. Balashov, V. A. Krylov,
   Thin Solid Films {\bf 239}, 127 (1994).
\bibitem{opps} S. Opps, B. Yang, C. Gray, D. Sullivan, to be published.
\bibitem{swanson}
  D. R. Swanson, R. J. Hardy, C. J. Eckhardt, 
    J. Chem. Phys. {\bf 99}, 8194 (1993).
  M. D. Gibson, D. R. Swanson, C. J.  Eckhardt, X. C. Zeng, 
    J. Chem. Phys. {\bf 106}, 1961 (1997).
\bibitem{barton} S. W. Barton, A. Goudot, O. Bouloussa, F. Rondelez,
  B. Lin, F. Novak, A. Acero, S.A. Rice, J. Chem. Phys. {\bf 96}, 1343 (1992).
\bibitem{me1} F. Schmid, M. Schick, J. Chem. Phys. {\bf 102}, 2080 (1995);
  F. Schmid, Phys. Rev. E {\bf 55}, 5574 (1997).
\bibitem{pink}
 D. A. Pink, T. J. Green, D. Chapman, Biochemistry {\bf 19}, 349 (1980);
 A.  Caill\'e, D.  Pink, F. de Verteuil, M. Zuckermann,
   Can. J. de Physique {\bf 58}, 581 (1980).
\bibitem{bernd-review}
  B. Dammann, H. C.  Fogedby, J. H. Ipsen, C. Jeppesen, K. Jorgensen,
    O. G. Mouritsen, J. Risbo, M. C. Sabra, M. M. Sperotto, M. J. Zuckermann,
    Handbook of nonmedical applications of liposomes, Vol. 1, pp. 85
    D. D. Lasic, Y. Barenholz eds., (CRC press, 1995).
\bibitem{szleifer}
  A. Ben-Shaul, I. Szleifer, W.M. Gelbart,
  J. Chem. Phys. {\bf 83}, 3597 (1985); 3612 (1985).
  I. Szleifer, A. Ben-Shaul, W.M. Gelbart,
  J. Chem. Phys. {\bf 85}, 5345 (1986).
\bibitem{rice-lat}
 J. Harris, S. A. Rice, J. Chem. Phys. {\bf 88}, 1298 (1987);
 S. Shin, Z. Wang, S.A. Rice,
  J. Chem. Phys. {\bf 92}, 1427 (1989); {\bf 93}, 5247 (1990).
\bibitem{kolinski}
  M. Milik, J. Skolnick, A. Kolinski, J. Phys. Chem. {\bf 96}, 4015 (1992);
  Y.K. Levine, A. Kolinski, J. Skolnick,
    J. Chem. Phys. {\bf 98}, 7581 (1993);
\bibitem{stettin}
 H. Stettin, H. J. M\"ogel, R. Friedemann,
    Ber. Bunsenges. Phys. Chem. {\bf 97}, 44 (1993);
 H. Stettin, H. J. M\"ogel, C. M. Care, 
    Ber. Bunsenges. Phys. Chem. {\bf 100}, 20 (1996).
\bibitem{haas1}
  F. M. Haas, P.-Y. Lai, K. Binder, 
     Makromol. Chem. Theory Simul. {\bf 2}, 889 (1993).
\bibitem{haas2}
  R. Hilfer, F. M. Haas, K. Binder, Nuovo Cimento {\bf 16}, 1297 (1994);
  F.M. Haas, R. Hilfer, K. Binder, J. Chem. Phys. {\bf 102}, 2960 (1995).
  F. M. Haas, R. Hilfer, K Binder, J. Phys. Chem. {\bf 100}, 15290 (1996).
\bibitem{haas3}
  F. M. Haas, R. Hilfer, J. Chem. Phys. {\bf 105}, 3859 (1996).
\bibitem{christoph1}
  F. Schmid, C. Stadler, H. Lange,
    Computer Simulations in Condensed Matter Vol. 10, pp. 37, 
      D. P. Landau, K. K. Mon, B. Sch\"uttler eds. (Springer, 1997).
\bibitem{harald-dipl} H. Lange, Diplomarbeit Universit\"at Mainz, 1996.
\bibitem{christoph-diss} C. Stadler, Dissertation Universit\"at Mainz, 1998.
\bibitem{me2} F. Schmid, H. Lange, J. Chem. Phys. {\bf 106}, 3757 (1997).
\bibitem{teer} B. Fischer, E. Teer, C. M. Knobler,
    J. Chem. Phys. {\bf 103}, 2365 (1995);
  E. Teer, C. M. Knobler, C. Lautz, S. Wurlitzer, J. Kildae, T. M. Fischer,
    J. Chem. Phys. {\bf 106}, 1913 (1997).
\bibitem{rigby}
  D.J. Rigby, R.J. Roe, J. Chem. Phys. {\bf 87}, 7285 (1987).
\bibitem{christoph2}
  C. Stadler, F. Schmid, in preparation.
\bibitem{allen-tildesley}
  M. P. Allen, D. J. Tildesley, {\em Computer Simulation of Liquids},
   (Oxford University Press, 1987).
\bibitem{cb}
  J. I. Siepmann, D. Frenkel, Mol. Phys. {\bf 75}, 59 (1992).
\bibitem{halperin-nelson}
  B. I. Halperin, D. R. Nelson, Phys. Rev. Lett. {\bf 41}, 121 (1978).
\bibitem{shih}
  M. C. Shih, T. M. Bohanon, J. M. Mikrut, P. Zschack, P. Dutta,
    J. Chem. Phys. {\bf 96}, 1556 (1991).
\bibitem{nigel}
  N. B. Wilding, A. D. Bruce, J. Phys. Cond. Matt. {\bf 4}, 3087 (1992).
\bibitem{footnote}
 In fact, these studies were done for a slightly larger head size:
 $\sigma_H = 1.2 \sigma$. For the liquid-gas transition, this should 
 however not make much difference.
\bibitem{footnote2} The fact that the head groups have been modelled as purely 
 repulsive units rather than Lennard-Jones beads does probably not make much
 difference. Adding an attractive part to the head interactions would certainly
 increase the probability of finding a liquid-gas transition, but presumably 
 not affect the condensed phases very much. The main effect of the head beads 
 in the condensed region is to put a restriction on the minimum grafting 
 distance between molecules.
\end{thebibliography}
\end{document}